%% file: ms.tex
\pdfoutput=1
\documentclass{article}

\usepackage[english]{babel}
\usepackage[utf8]{inputenc}
\usepackage{color}

\usepackage{graphicx} %% figures
\usepackage{amsmath} %% maths
\usepackage{amssymb} %% maths
\usepackage{amsthm} %% maths
\usepackage{tabularx} %% tableaux largeur automatique
\usepackage{siunitx} %% unites
\usepackage{algorithm} %% algos
\usepackage[noend]{algpseudocode} %% algos
\usepackage{caption} %% captions et subcaptions (a) (b) (c) etc.
\usepackage{subcaption} %% captions et subcaptions (a) (b) (c) etc.
\usepackage{filecontents} %% bibliographie dans le fichier tex

\usepackage{authblk}

\everymath{\displaystyle}
\graphicspath{{img/}}

\title{ A new front-tracking Lagrangian model for the modeling of dynamic and post-dynamic recrystallization }% Force line breaks with \\
\author[1]{Sebastian~Florez\thanks{corresponding author}}
\author[1]{Karen~Alvarado}
\author[1]{Marc~Bernacki}

\affil[1]{Mines-ParisTech, PSL-Research University, CEMEF – Centre de mise en forme des mat\'{e}riaux, CNRS UMR 7635, CS 10207 rue Claude Daunesse, 06904 Sophia Antipolis Cedex, France}%

\begin{document}
\maketitle
%\begin{frontmatter}

\section*{abstract}

A new method for the simulation of evolving multi-domains problems has been introduced in previous works (RealIMotion), Florez et al. (2020) and further developed in parallel in the context of isotropic Grain Growth (GG) with no consideration for the effects of the Stored Energy (SE) due to dislocations. The methodology consists in a new front-tracking approach where one of the originality is that not only interfaces between grains are discretized but their bulks are also meshed and topological changes of the domains are driven by selective local remeshing operations performed on the Finite Element (FE) mesh. In this article, further developments and studies of the model will be presented, mainly on the development of a model taking into account grain boundary migration by (GBM) SE. Further developments for the nucleation of new grains will be presented, allowing to model Dynamic Recrystallization (DRX) and Post-Dynamic Recrystallization (PDRX) phenomena. The accuracy and the performance of the numerical algorithms have been proven to be very promising in Florez et al. (2020). Here the results for multiple test cases will be given in order to validate the accuracy of the model taking into account GG and SE. The computational performance will be evaluated for the DRX and PDRX mechanisms and compared to a classical Finite Element (FE) framework using a Level-Set (LS) formulation.

\include{Introduction}

\include{NumericalMethod}

\include{NumericalResults}

\include{Conclusions}

\bibliography{ms}

\end{document}

%% file: Introduction.tex
\section{Introduction}

The modeling, at the mesoscopic scale, of Grain Growth (GG) and recrystallization (ReX) in polycrystalline materials during thermal and mechanical treatments has been the focus of numerous studies in the last decades. Indeed, mechanical and functional properties of metals are strongly related to their microstructures which are themselves inherited from thermal and mechanical processing.\\%The reason behind this tendency is the clear importance of the microstructural estate of materials regarding their final mechanical properties, hence making predictions on the grain size and grain morphology via numerical simulations a very important field of research.\\

When looking to the so-called full-field (FF) methods, based on a full description of the microstructure topology and modeling of grain boundary migration (GBM) at mesoscopic scale, main numerical frameworks involve: Monte Carlo (MC) \cite{Rollett1989, Rollett2001}, Cellular Automata (CA) \cite{Raabe2002, BarralesMora2008, Rauch2015, Madej2018}, Multi Phase-Field (MPF) \cite{Steinbach1996, Moelans2008, Krill2002, Kim2014}, Vertex/Front-Tracking \cite{kawasaki1989, weygand1998, Lepinoux2010, BarralesMora2010, Mellbin2015} or Level-Set (LS) \cite{Merriman1994, Bernacki2008,Cruz-Fabiano2014,maire2016} models. These numerical methods are developed by many researchers \cite{Humphreys2017}. All the mentioned methods have, of course, their own strengths and weaknesses \cite{Humphreys2017, Bernacki2019}.\\

When large deformation have to be considered (common in metal forming context), LS or MPF approaches in context of unstructured FE mesh and FE remeshing strategies remain the main powerful and generic approaches but with a strong limitation in terms of computational cost.\\

In this context vertex and front tracking approaches appear as interesting candidates. An explicit description of the interfaces is considered and GBM is imposed at each increment by computing the velocity of the nodes describing the interfaces. While having a deterministic resolution (solving of partial differential equation - PDE), this methodology is very efficient. However, the implementation of the topological events is not straightforward and the fact to not describe the bulk of the grains could be limiting for some metallurgical mechanisms such as appearance of new grains (nucleation) or substructures inside the grains.\\

Previous works dedicated to the creation of an improved front-tracking method, solving these weaknesses, have been published in previous articles \cite{Florez2020b, Florez2020c}. The model denominated TOpological REmeshing in lAgrangian framework for Large interface MOTION (ToRealMotion, hereafter TRM) maintains the interior of grains meshed, handling with relative ease the topological changes of the grain microstructure and allowing the treatment of in-grain operations and at a higher computational performance than classical FE-LS models for the same accuracy. The objective of the present article is then to adapt the TRM model to handle DRX and PDRX phenomena.
 %operations and to develop a GG model capable to take into account the impact of the dislocation density in the kinetics of grain boundaries in the form of Stored Energy (SE).

%% file: NumericalMethod.tex
\section{The TRM model : Isotropic Grain Growth context} \label{sec:theTRMmodel}

The TRM model has been presented in a previous work in \cite{Florez2020b}, then adapted to a parallel computational environment in \cite{Florez2020c}. This model uses the logic behind front-tracking methods where the discretization of interfaces is the minimal topological information allowing to model 2D-GBM. The TRM model goes a step further by implementing also a discretization of the interior of the grains in the form of simplexes to allow the interaction of the grain boundaries with the bulk of the grains and preventing inconsistencies of the physical domain such as the overlapping of regions. The data structure of the TRM model is then built on top of a mesh with element and nodes, enabling also the possibility to compute FE problems on it. This data structure defines geometrical entities such as \emph{points}, \emph{lines} and \emph{surfaces} by grouping sets of nodes and elements: each \emph{point} regroups a P-Node\footnote{Which defines a node of the mesh with a topology degree equal to 0 on the microstructural framework, hence a multiple junction.} and a set of connections to other \emph{points} and \emph{lines}. Each \emph{line} is defined by an ordered set of L-Nodes\footnote{Nodes with a topology degree equal to 1, or a node belonging to a simple grain boundary in the microstructure.}, an initial \emph{point} and a final \emph{point}. Finally, \emph{surfaces} are defined as a set of S-Nodes\footnote{Nodes with a topology degree equal to 2, or nodes in the bulk of grains with no connection to the S-Nodes on other grains.}, a set of elements and a set of delimiting \emph{lines} and \emph{points}.\\

This data structure can be constructed by performing some preprocessing steps, in \cite{Florez2020b}, a preprocessor able to transform MPF or LS data to the data structure of the TRM model has also been introduced. This preprocessor is based on the works presented in \cite{Shakoor2017ijnme} and in \cite{Florez2020}, where a remeshing procedure transforms a typical LS configuration (a FE mesh with grain boundaries being defined by the interpolated zero iso-value of several LS fields as in \cite{Scholtes2015, Scholtes2016}) into a body-fitted mesh (where the interpolated zero iso-value coincide with some nodes of the FE mesh) with the help of a  \emph{joining and fitting algorithm}, expliciting the nodes on the ``front" to track (as in front-tracking methods). Subsequently on the preprocessing step, a classification is made for the nodes of the mesh based on their topological representation on the microstructural space. Finally, a geometric identification algorithm is performed to build the geometrical entities mentioned before. The reader is referred to \cite{Florez2020b} for a complete description of the reconstruction process.\\

After the preprocessor, the data in the form of LS fields in no longer needed as the microstructure is now defined by the identified geometric entities. The classification of geometric entities is also helpful when computing geometric properties, the area of surfaces can be computed by adding the contribution of each element of the grain while the curvature $\kappa$ and normal $\vec{n}$ of interfaces can be obtained by approximating the interface with a high order mathematical form (higher than the linear discretization of the domain) such as a least square approximation or with piece-wise polynomials such as natural parametric splines. We have opted to use the latter in order to obtain such geometrical quantities.\\

Once the data structure is defined, the \emph{physical} mechanism can be simulated. This physical mechanism represents how the different geometries are supposed to evolve based on their current state. The TRM model has been developed to move the different nodes of the mesh based on a user defined velocity field $\vec{v}$ and a time step $dt$. Once a velocity is defined a new position for each node $N_i$ on the mesh can be obtained as:\\

\begin{equation}
\label{Eq:lagrangianDisplacement}
\centering
\vec{u_i}=\vec{u_{i}^0} + \vec{v_i} \cdot dt,
\end{equation}

where $\vec{u_{i}^0}$ is the current position of the node $N_i$. 

Here, each node displacement can potentially produce an overlap \footnote{An overlap in a mesh is produced when an element is partially or completely superposed by another element hence disrupting the 1:1 mapping of the numerical domain to the physical domain, such a mesh can not be used in a Finite Element resolution.} of some of the elements attached to the node. The TRM model hence ensures the local conformity of the mesh by means of a ``local-iteratively movement-halving" that finds iteratively the approximated maximal displacement that a node is able to make in the direction of the velocity $v_i$ when an overlap takes place. This procedure ensures at all times that both, the mesh and the microstructural domain are valid.\\

Once several steps of Lagrangian movement are performed, it is highly probable that the quality of the mesh become too poor to continue with the GBM, this is why the TRM model implements a particular remeshing procedure strongly influenced by the works in \cite{Compere2008, Compere2009}, that improves the mesh quality and allows topological events such as grain disappearance. This remeshing procedure must be adapted to the data structure previously defined and it has to ensure that the final mesh after remeshing is valid to be used by the TRM model. Some of the geometrical entities must adapt their sets to take into account the changes made on the mesh, to do this, several local \emph{selective}\footnote{The word selective denotes a variation of the original remeshing operations when performed over the data structure of the TRM model, as each remeshing operation will be performed differently over nodes with different topology (P-Node, L-Node and S-Node).} remeshing operators have been developed that allow to make changes on the mesh by maintaining a valid data structure: selective vertex smoothing, selective node collapsing, selective edge splitting, selective edge swapping and selective vertex gliding, see \cite{Florez2020b} for a complete description of these operators and the global remeshing procedure.

\subsection{Parallel Implementation}

\begin{figure}[!h]
\centering
\includegraphics[width=0.5\textwidth] {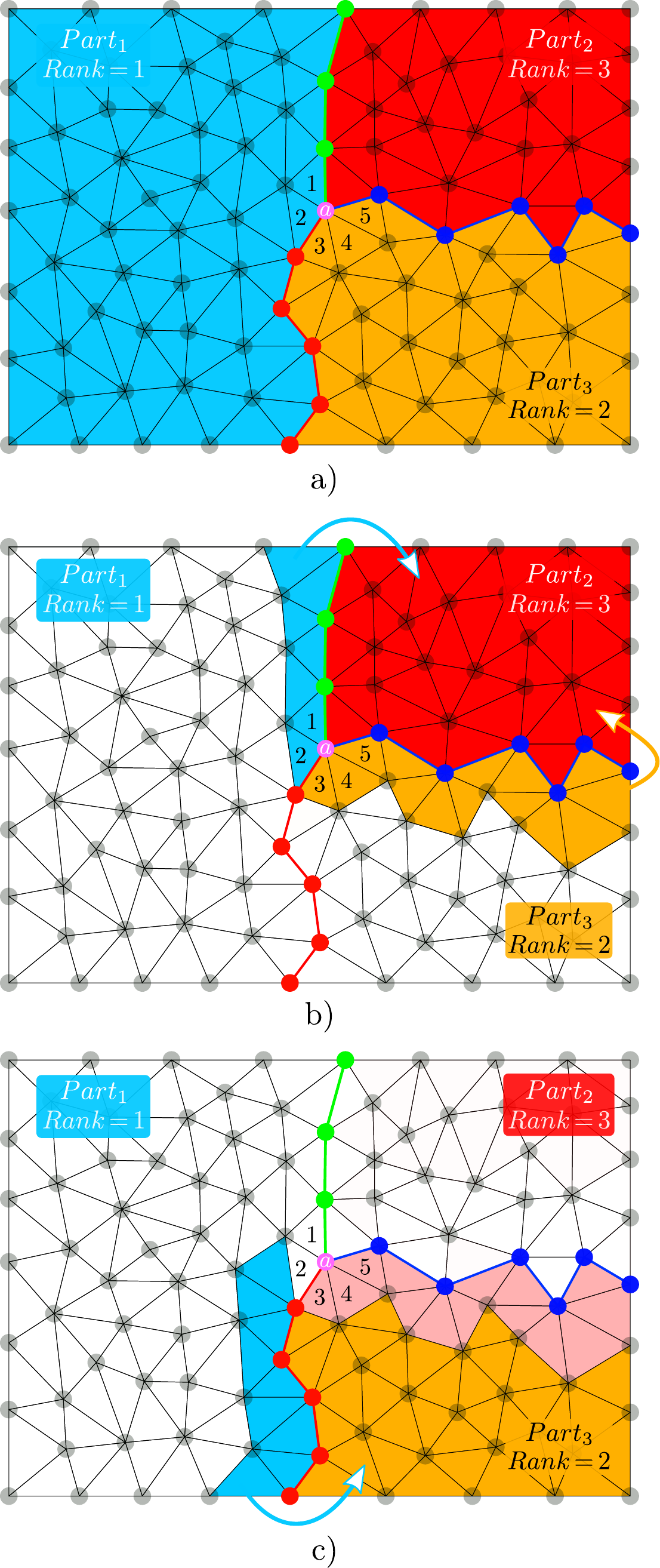}
\caption{ Example of the behavior of the Unidirectional selection element algorithm. a) initial state with three parts, the name and the rank of each partition is displayed. b) Selected elements to be sent to $Part_2$, elements 3, 4 and 5 appear only to be sent to $Part_2$, and not to $Part_1$. Elements 1 and 2 appear initially on the list to be sent to $Part_2$ and $Part_3$ but they are filtered in the last part of the algorithm, as the higher rank of the nodes of these elements belongs to $Part_2$. c) Selected elements to be sent to $Part_3$, the intersection of elements from $Part_1$ to be sent to $Part_2$ and $Part_3$ is empty \cite{Florez2020c}.}
\label{fig:UnidirectionalSelection}
\end{figure}

As presented in \cite{Florez2020c}, a \emph{distributed memory}\footnote{Approach where each processor interacts with its own independent memory location.} system, was chosen for our model, allowing the computation of simulation across single CPUs and over a computational cluster where multiple CPUs are connected over the local network and not over a single mother board. For this implementation, the standard communication protocol Message Passing Interface (MPI) \cite{Walker1992} was used and the free library METIS \cite{Karypis1998} was used in order to obtain the initial partitioning of the domain. Additionally, multiple new functions were added to the original sequential approach in order to make it work in parallel. Among these functions, the implementation of a repartitioning algorithm was essential to obtain a re-equilibrium of charges as well as a coherent remeshing approach. This repartitioning algorithm is based in a user-defined ranking system, where a unique rank is attributed to each processor (in \cite{Florez2020c} the number of elements was used to define the ranks), then a \emph{Unidirectional Element Sending} algorithm was developed to exchange layers of elements between the partitions as illustrated in Fig. \ref{fig:UnidirectionalSelection}.\\

The remeshing strategy was developed based on the blocking of the boundaries between partitions, meaning that any application of a remeshing operator that changes the boundaries between partition is discarded by default. Nor the creation or the deletion of nodes (and edges) is allowed at the boundary, also \emph{vertex smoothing is not allowed} for the Shared-Nodes\footnote{Nodes present in the memory of multiple processor at the same time, these nodes reside at the boundaries between partitions.}. Finally, in order to obtain a complete remeshing over the hole domain, the \emph{Unidirectional Element Sending} algorithm, ensures the motion of the boundaries between partitions, hence unblocking those edges before the element scattering. This is very convenient as not a lot of changes are necessary over the original implementation of the remeshing operations but inconvenient as the remeshing must be performed two times in a single increment.\\

Some other algorithms are necessary for the complete parallel implementation, these algorithms handle the identification and reconstruction of geometric entities across the segmented domain as well as the computation of geometric properties and the Lagrangian movement of the Shared-Nodes, see \cite{Florez2020c} for a complete description.

\section{Grain boundary migration under capillarity and SE driving pressures}

The simulation of microstructural evolutions are given by the addition of complex and different phenomena as GG \cite{Bernacki2008,Cruz-Fabiano2014, Hallberg2014,maire2016,furstoss2018}, Recrystallization (ReX)  \cite{Bernacki2008, Bernacki2009, Bernacki2011, Scholtes2015, Scholtes2016, Maire2017, Hallberg2013}  or Zener Pinning (ZP) \cite{Weygand1999, Couturier2003, Couturier2004, Couturier2005, Agnoli2015}. In \cite{Florez2020b}, isotropic GG with no influence of SE was used to compare the TRM model to other approaches (LS-FE \cite{Bernacki2011,Cruz-Fabiano2014,maire2016}), the base model used to represent this phenomenon is commonly known as migration by curvature flow. The velocity $\vec{v}$ at every point on the interfaces can be approximated by the following equation:

\begin{equation}
\label{Eq:VelocityEquationc}
\centering
\vec{v_c}=-M \gamma\kappa \vec{n},
\end{equation}

where $M$ is the mobility of the interface, $\gamma$ the grain boundary energy,  $\kappa$ the local magnitude of the curvature in 2D and $\vec{n}$ the unit normal to the grain interface pointing to its convex direction. In an isotropic context as considered here, the terms $M$ and $\gamma$ are supposed as invariant in space.\\

%Of course, capillarity is not the only mechanism present during GG, other properties such as the dislocation density present on the crysalline lattice of a grain can determine how their boundaries may react to their viccinity. This property of a grain is commonly known as its Stored Energy.
Of course when post-dynamic phenomena such as Static ReX (SRX) or Meta-Dynamic ReX (MDRX) are considered, the SE will act as another driving pressure of the GBM. Note that the SE within a grain can be variant, as there could be regions on the grain that have accumulated more or less dislocations during the considered thermomechanical treatment (TMT).
%however, the influence of a gradient on the dislocation of a single grain on the kinetics of their interfaces has not been widely studied on the litterature at the mesoscopic scale. Currently, Methods such as \emph{Crystal Plasticity} (CP) \cite{davidRuiz Articles} are able to obtain and use such gradients into a simulation at a very high computational cost.  Moreover previous works in the context of LS-FE formulations \cite{Ilin2018} have studied this phenomenon, concluding that inter-granular gradients on the stored energy could indeed have a big impact on the morphology of grains and that simulations taking into account such variations were more in accordance with the experimental results than simulations using a constant value of stored energy per grain. This is a big setback for methods such Vertex and front-tracking models which do not maintain a discretization of the interior of grains making it very difficult to use spacial gradients, to the knowledge of the authors there have been no attempts to circumvent this issue in the context of front-tracking models.
%The influence of a gradient on the dislocation of a single grain on the kinetics of their boundaries is out of the scope of this article, this is why in section \ref{sec:results} we will use a constant value to express the SE value within a grain, nonetheless the approach presented in this article to model GG by SE can be used in the context where a gradient of SE is present and will be studied in future works.
At the mesoscopic scale, the SE can be discussed following different hypotheses. Crystal plasticity calculations and EBSD experimental data can bring dislocation density field and so SE field with fine precision until intragranular heterogeneities. While this information is directly usable in pixel/voxel based stochastic approaches such as MS or CA methodologies, generally it is homogenized by considering constant value per grain in deterministic front-capturing (MPF, LS) and front-tracking approaches. If this choice seems quite natural for phenomena where stored energy gradients and nucleation of new grains are mainly focused on GB like for discontinuous DRX (DDRX), it could be a strong assumption for phenomena where the substructure evolution is important, like for continuous DRX (CDRX). This aspect was for example studied in \cite{Ilin2018} in context of SRX with a FE-LS numerical framework. It was concluded that intragranular gradients on the stored energy could indeed have a big impact on the grain morphology and that simulations taken into account such variations were more in accordance with experimental observations than simulation using a constant value of stored energy, but with an important numerical cost as the FE mesh must be then adapted at the intragranular heterogeneities scale. In the following, a constant homogenized energy per grain is assumed. Nonetheless, the approach presented in this article to model GG with a stored energy field can be used in the context of a heterogeneous intragranular energy field, this aspect will be investigated in a forthcoming publication.

Thus, here SE can act on the displacement of the interface by considering the difference of SE at both sides of the interface. We will adopt a slightly modified methodology to the one presented in \cite{Bernacki2009} to quantify it: 

\begin{equation}
\label{Eq:VelocityEquatione}
\centering
\vec{v_e}=-M \delta_{(\dot{\epsilon})} [E]_{ij} \vec{n},
\end{equation}

where the term $[E]_{ij}$ defines the difference of stored energy $E$ between the grains $i$ and $j$ ($E_i-E_j$), the term $\delta_{(\dot{\epsilon})}$ is a mobility coupling factor whose nature is explained in \cite{Maire2017} appendix c\footnote{A mobility coupling factor function of the effective strain rate $\dot{\epsilon}$ with $\delta_{(\dot{\epsilon})}=1$ when $\dot{\epsilon}=0$.} and where the direction of the unit normal $\vec{n}$ sets, for a given node of the interface, the order of the indices as: first the index $i$ and then $j$.
% This means that if the value of $E_i$ is inferior to $E_j$ the velocity $\vec{v_e}$ will be pointed on the same direction as $\vec{n}$, on the other hand, if the value of $E_j$ is inferior, the velocity $\vec{v_e}$ will be pointed contrary to $\vec{n}$.
Note that this definition holds even if the direction of $\vec{n}$ is ambiguous (in the case of a flat interface with no convex side) as the direction of the velocity $\vec{v_e}$ will be then pointed, in all cases, from the lower to the higher value of stored energy no matter what the direction of $\vec{n}$ is. Moreover the value of stored energy can be computed using the equation:

\begin{equation}
\label{Eq:StoredEnergyFromRho}
\centering
E=\dfrac{1}{2}\mu b^2 \rho,
\end{equation}

where $b$ corresponds to the norm of the Burgers vector and $\mu$ correspond  to the elastic shear modulus of the material.

Finally, the contribution of driving pressures due to SE and capillarity can be accounted by linearly adding the two velocities as in \cite{Bernacki2009, Scholtes2016}:

\begin{equation}
\label{Eq:VelocityEquation}
\centering
\vec{v}=-M ( \delta_{(\dot{\epsilon})} [E]_{ij} \vec{n} +\gamma\kappa \vec{n}),
\end{equation}

where $\vec{v}$ denotes the final velocity of the interface during GBM when SE effects are included.

\subsection{Velocity at Multiple Junctions} \label{sec:multJunctVelocity}

Equation \ref{Eq:VelocityEquationc} can only be used in a one-boundary problem, as in a more general context, the presence of multiple junctions (the intersection points of more than 3 interfaces) makes it impossible to compute a curvature $\kappa$ or a normal $\vec{n}$ at these points. As explained in previous works \cite{Florez2020b, Florez2020c}, we have used an alternative methodology to compute the velocity due to capillarity at multiple points: Model II of \cite{kawasaki1989}, where the product $\kappa \vec{n}$ is directly obtained from an approximation of the free energy equation of the hole system in a vertex context.\\

\begin{figure}[!h]
\centering
\includegraphics[width=0.70\textwidth] {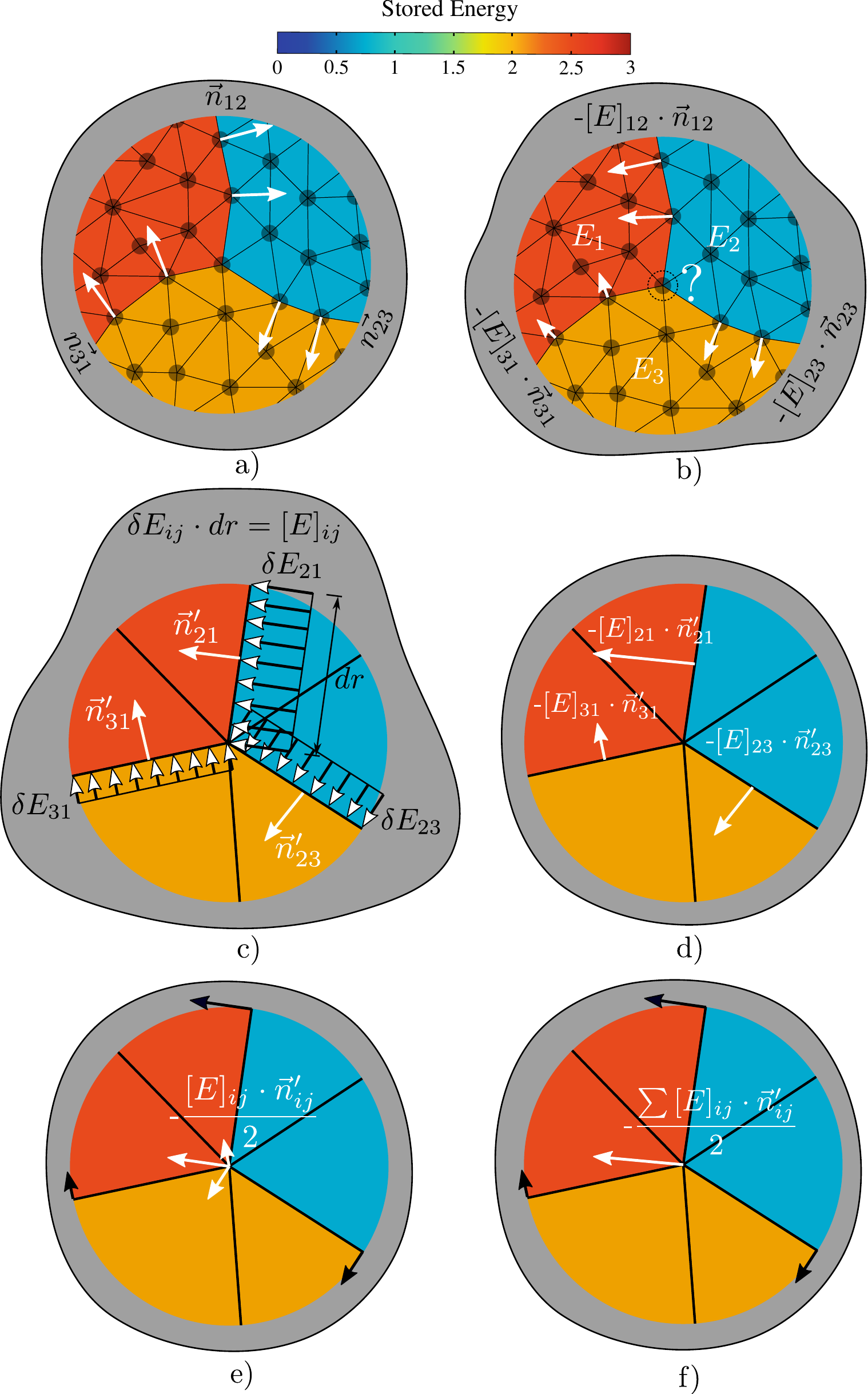}
\caption{Graphical demonstration of the obtention of Eq.~\ref{Eq:VelocityEquationeMP}, a) typical triple junction configuration with values of SE homogenized on each grain and the normal vectors $n$ computed at the nodes of the interfaces pointing to their convex side, b) computation of the term $-[E]_{ij} \cdot \vec{n_{ij}}$ for each node of the interface except for the node at the triple junction, c) definition of the same configuration as in a) but in a differential portion of radius $dr$, d) the resultant driving forces are applied at the center of the segments on the differential portion, e) and f) the driving forces are distributed at the ends of each segment and an expression can be formulated at the triple junction for its resultant driving force.}
\label{fig:StoredEnergyComp}
\end{figure}

Similarly, Eq. \ref{Eq:VelocityEquatione} only holds in a one-boundary problem as neither the value of $[E]_{ij}$ nor the value of $\vec{n}$ can be obtained at these points. To solve this, a different approach has been developed to compute a ``resultant" velocity due to store energy $\vec{v_e}$ at multiple junctions. This approach is illustrated in Fig. \ref{fig:StoredEnergyComp} where for the sake of clarity, the value of $M$ has been held constant and equal to $1$. Fig. \ref{fig:StoredEnergyComp} a) shows a typical configuration where the boundaries of three grains converge to a single point, each grain $i$ has its own stored energy $E_i$ where $E_1>E_3>E_2$. The values of the velocity for each normal boundary have been computed with Eq. \ref{Eq:VelocityEquatione} and are shown as white arrows for each node in the boundary of Fig. \ref{fig:StoredEnergyComp} b), here the index on the normal $\vec{n_{ij}}$ term are only representative of their direction and serve to set the indices of each $[E]_{ij}$ terms, these expressions do not follow the Einstein notation summation laws, all summations will be represented by the conventional $\Sigma$ operator.\\

If a portion of differential size $dr$ centered at the multiple point is evaluated (see Fig. \ref{fig:StoredEnergyComp} c)) the boundaries between grains will appear as flat, here the difference on the stored energy can be seen as a distributed difference of potential $[E]_{ij}$ applied on the length of the grain boundary of size $dr$ (analog to a given pressure acting as a resultant force on a given interface). A normal $\vec{n}'_{ij}$ can be obtained and used to compute a velocity of each boundary  (\ref{fig:StoredEnergyComp} d)) applied at its center. Note that the direction of $\vec{n}'_{ij}$ can be chosen ambiguously on this linear segment, however, as mentioned before, an eventual ambiguity on the direction of $\vec{n}$ do not represent an ambiguity on the term $-[E]_{ij} \cdot \vec{n_{ij}}$ as $[E]_{ij} \cdot \vec{n_{ij}} = [E]_{ji} \cdot \vec{n_{ji}}$ with $[E]_{ji}=-[E]_{ij}$ and $\vec{n_{ji}}=-\vec{n_{ij}}$. These velocities can be divided and applied at the ends of each boundary and finally added at the junction point (\ref{fig:StoredEnergyComp} e) and f) respectively) to obtain a valid velocity vector field at multiple junctions. The expression on Fig. \ref{fig:StoredEnergyComp} f) can be extended to the case where the values of $M$ are neither constant nor equal to 1:

\begin{equation}
\label{Eq:VelocityEquationeMP}
\centering
\vec{v_e}=\dfrac{-\Sigma M \delta_{(\dot{\epsilon})} [E]_{ij} \vec{n}}{2},
\end{equation}

of course, this expression can be also used in cases of multiple junctions of any order, where more than three interfaces meet. Eq. \ref{Eq:VelocityEquationeMP} will be used to compute the value of $v_e$ at multiple junctions as an approximation to the yet unknown behavior of such configurations under the influence of stored energy in a transient state.

%Finally, note that this equation also holds if a differential portion centered at normal grain boundary (not a multiple junction) is analysed, as in this case, only two sides of the boundary will contribute to the velocity at the center, hence adding two times the same term $M [E]_{ij} \vec{n}$ (if the difference of stored energy is constant in the tangent direction of the interfarce), obtaining the same expression as in equation \ref{Eq:VelocityEquatione} (provided that the normal direction is also constant in the differential portion). The same analysis can be held for non-constant values of $[E]_{ij}$ in the tangent direction, i.e. if the value of stored energy is constant per element: Each interface (edge in the mesh) is able to compute a $[E]_{ij}$, then each node of the boundary will use its two attached boundary edges to compute its velocity $v_e$ with \ref{Eq:VelocityEquatione}.

\subsection{Topological changes: capillarity, stored energy}

Multiple changes on the topology of the microstructure occur during GG and ReX. In general, the topological changes during GG are given by the disappearance of grains: on a shrinking grain, each of their boundaries evolve until they \emph{collapse} to multiple junctions. Eventually, all boundaries collapse to a single multiple junction and the domain occupied by the grain disappears. This behavior was implemented on the original TRM model presented in \cite{Florez2020b} by means of the application of the \emph{selective node collapse} operator, where some restrictions where made regarding the order of collapsing.
%: In this article, only GG by capillarity was taken into account. In fact, GG by capillarity is principally led by the motion of multiple junctions, consider the case presented in figure \ref{fig:TriplePointInitialAndSteady} corresponding to the evolution of an initial T-junction problem with 3 grains. Here the all initial grain boundaries (\ref{fig:TriplePointInitialAndSteady}a)) are flat, hence their velocity is 0, except for the multiple junction. Once the multiple junction moves, a curvature of two of their adjacent boundaries starts to form (\ref{fig:TriplePointInitialAndSteady}b)) producing a growing velocity. Finally the profile stabilizes obtaining a constant velocity for the two curved boundaries (\ref{fig:TriplePointInitialAndSteady}c)). This is a very common case in GG by capillarity that suggest that multiple points are responsible for the movement of the global microstructure when only capillarity is taken into account.

%\begin{figure} [h]
%	\centering
%	\includegraphics[width=0.9\textwidth]{img/TriplePoint.pdf}
%	\caption{\label{fig:TriplePointInitialAndSteady}T-Junction Case. left: initial state, center: transient state, right: steady state. }
%\end{figure}

\begin{figure}[!h]
\centering
\includegraphics[width=0.5\textwidth] {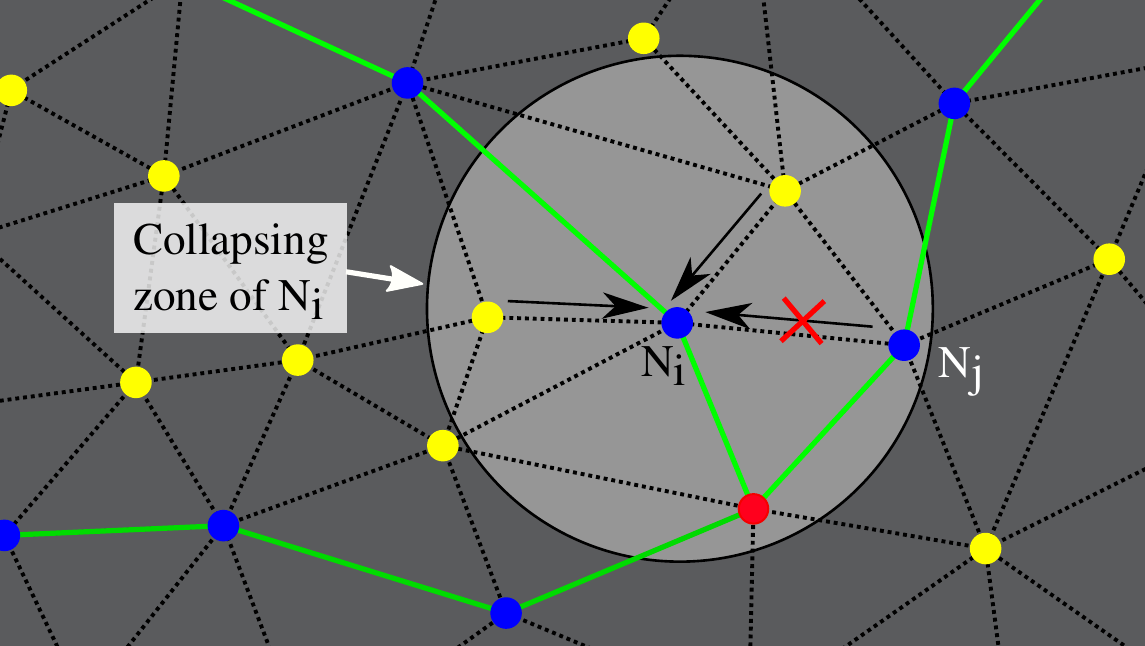}
\caption{Node Collapsing rule in GG by capillarity. Some nodes are within the collapsing zone of $N_i$: Two S-Nodes (yellow) will collapse, one L-Node (blue) cannot collapse and one P-Node (red) cannot collapse. $N_i$ cannot collapse P-Nodes (topological degree), $N_i$ can collapse L-Nodes but $N_j$ does not belong to the same line (i.e. they do not belong to the same grain boundary). \cite{Florez2020b}}
\label{fig:CollapseInterfaceCondition1}
\end{figure}

\begin{figure}[!h]
\centering
\includegraphics[width=0.5\textwidth] {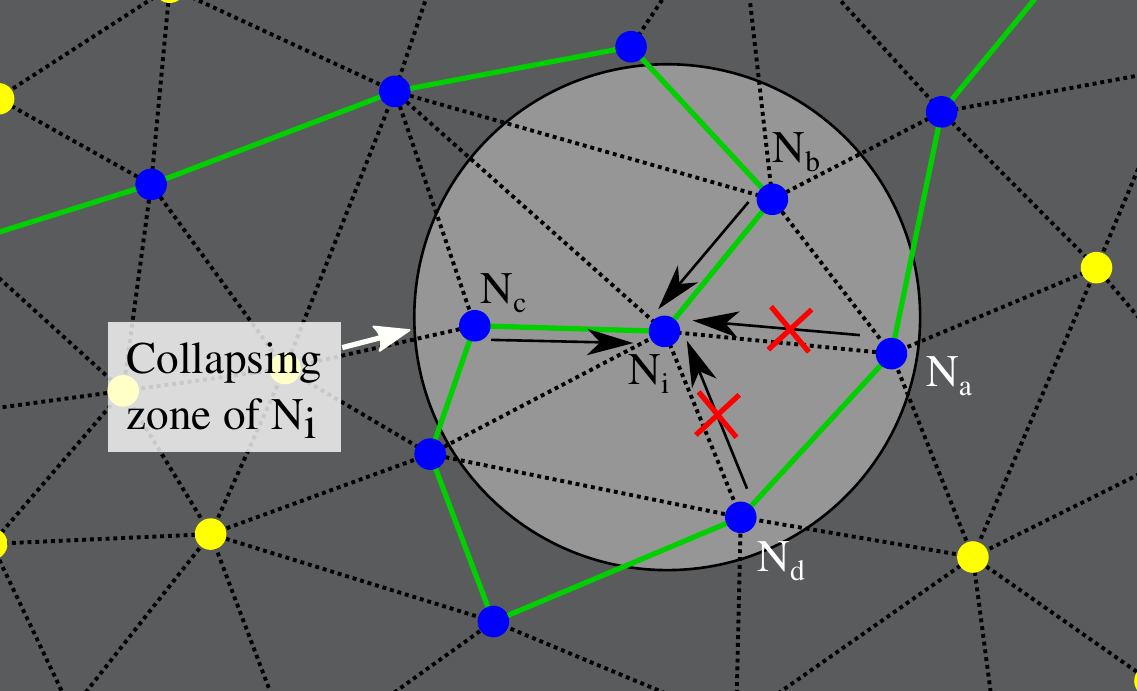}
\caption{Node Collapsing rule in GG by capillarity. Some nodes are within the collapsing zone of $N_i$: four L-Nodes (blue) $N_a$, $N_b$, $N_c$ and $N_d$. Only Nodes $N_b$ and $N_c$ can collapse as they are consecutive to $N_i$ within the same line. \cite{Florez2020b}}
\label{fig:CollapseInterfaceCondition2}
\end{figure}

In \cite{Florez2020b} we had opted to use this methodology to produce coherent topological changes on the microstructure, leading to a series of rules on the \emph{selective node collapse} operator (see section 2.4.1 of \cite{Florez2020b}). These rules gave to the P-Nodes a higher influence over other kind of nodes and prevented the collapsing of non-consecutive nodes as illustrated in Figures \ref{fig:CollapseInterfaceCondition1} and \ref{fig:CollapseInterfaceCondition2} respectively.

The implementation of such node collapsing strategy allows a high control over the order on which the topological changes occur, unfortunately this kind of reasoning can only be used on isotropic GG and can not be used when SE or spatial heterogeneities of the mobility/interface energy must be taken into account.\\

\begin{figure}[!h]
\centering
\includegraphics[width=1.0\textwidth] {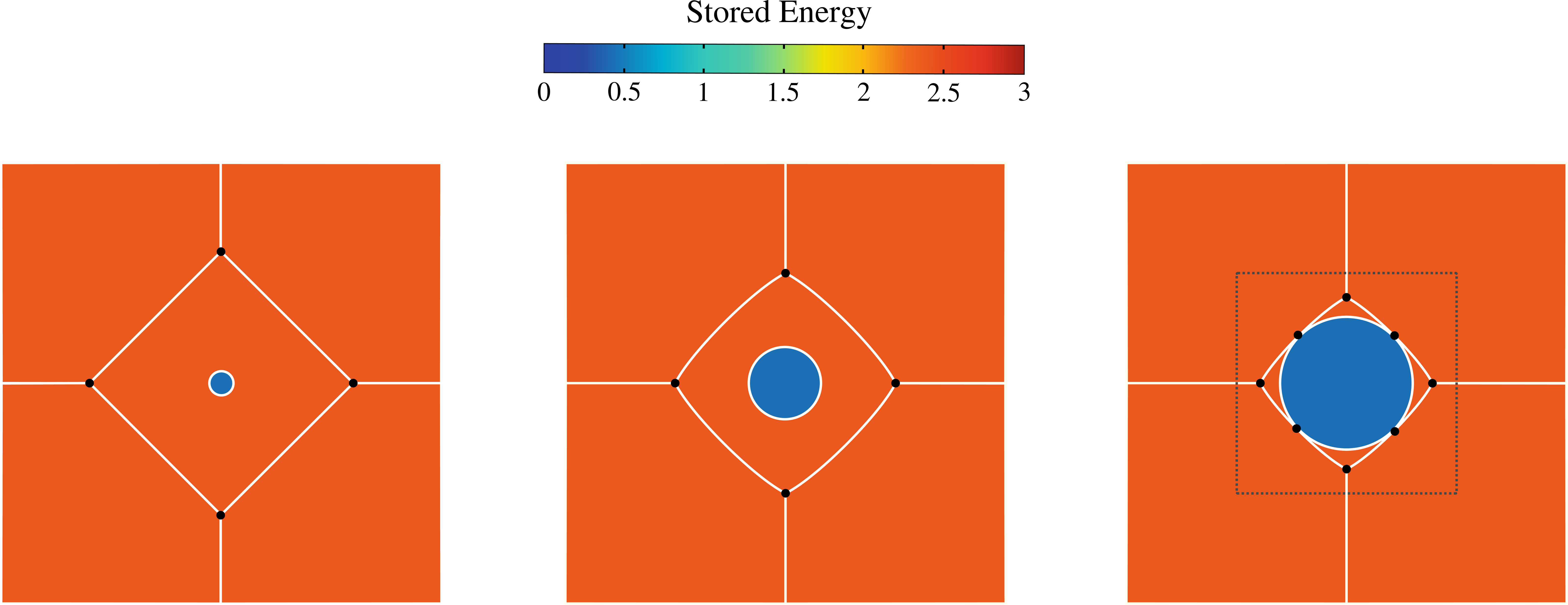}
\caption{Six grains with a specific SE balance, the circular grain in the middle grows due to its low SE compared to the SE of its surrounding grains. The initially squared grain shrinks by the combined effects of capillarity at their external boundaries and the surface taken away by the circular growing grain. Left: initial state, center: the circular grain grows, right: the boundary of the circular grain and the external boundaries of the initially square grain collide.}
\label{fig:StoredEnergySquareShrinkage}
\end{figure}

\begin{figure}[!h]
\centering
\includegraphics[width=1.0\textwidth] {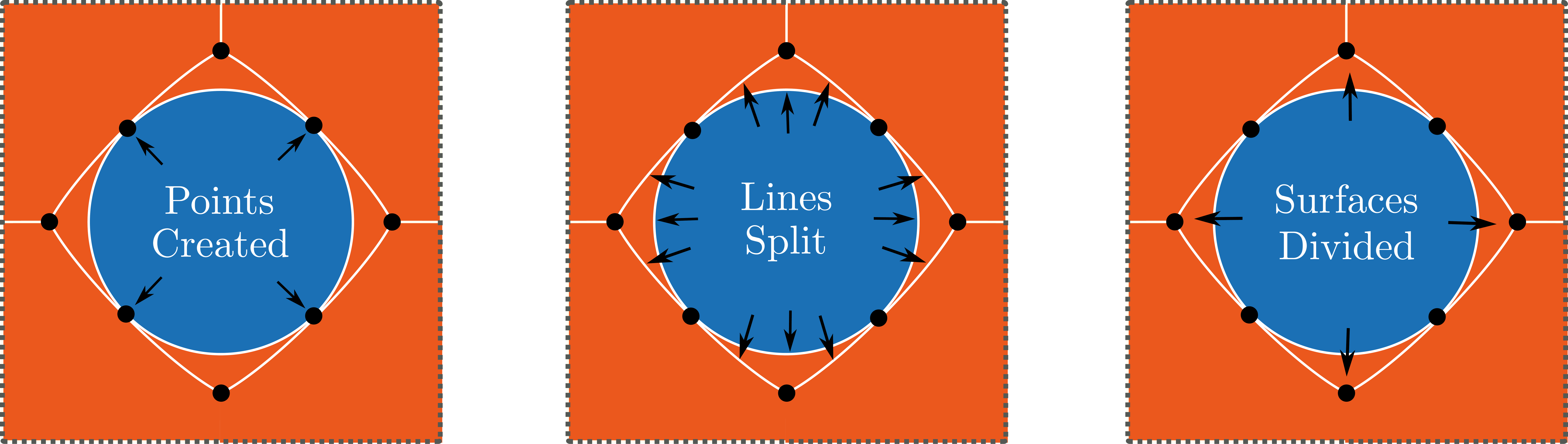}
\caption{Details of the final event of Fig. \ref{fig:StoredEnergySquareShrinkage}, highlighting the changes on the microstructure. Here, Points describe multiple junctions, Lines grain boundaries and Surfaces grains.}
\label{fig:StoredEnergySquareShrinkage_EventDetail}
\end{figure}

When considering SE, the kinetics of the GB are not only led by the movement of multiple points; flat surfaces can evolve with a given velocity and it is possible that the velocity of simple boundaries becomes much more important than the velocity of multiple junctions. Fig. \ref{fig:StoredEnergySquareShrinkage} illustrates this behavior with six grains with a specific SE state. The circular grain in the middle grows due to its low SE compared to the SE of its surrounding grains. The circular grain is indeed surrounded by an initially squared grain that starts shrinking by the combined effects of capillarity at their external boundaries and the surface taken away by the circular growing grain. Fig. \ref{fig:StoredEnergySquareShrinkage} (right), shows the moment when the boundary of the circular grain and the external boundaries of the initially square grain collide, unchaining a series of topological changes on the microstructure. These topological changes are illustrated in Fig. \ref{fig:StoredEnergySquareShrinkage_EventDetail}, where new multiple junctions (Points) appear, grain boundaries (Lines) are split and grains (Surfaces) are divided. \\

\begin{figure}[!h]
\centering
\includegraphics[width=1.0\textwidth] {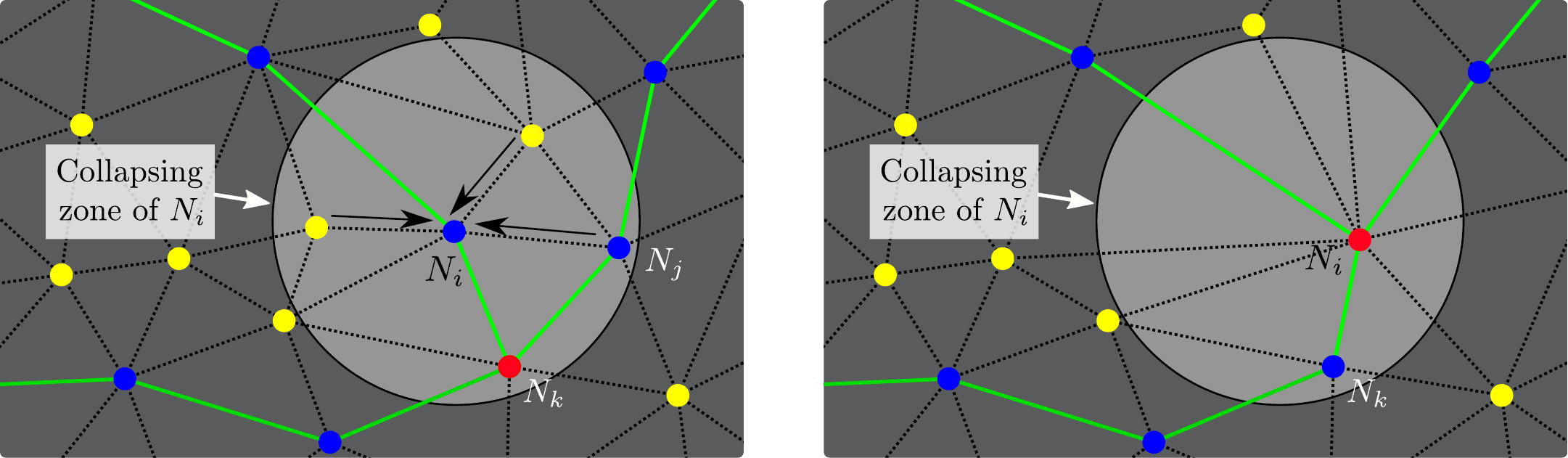}
\caption{Node Collapsing of Fig. \ref{fig:CollapseInterfaceCondition1} when allowing the collapse between non consecutive nodes, S-Nodes are displayed in yellow, L-Node in blue and P-Nodes in red. The collapsing of nodes $N_i$ and $N_j$ produces a new P-Node ($N_i$) while the pre-existent P-Node $N_k$ needs to be reclassified as a L-Node. Left: initial state, right: state after collapse. }
\label{fig:CollapseInterfaceCondition_Override_1}
\end{figure}

These several changes on the microstructure (contact of different grain boundaries in non-convex grains) can not be accomplished by the TRM model if the rules described above and illustrated by figures \ref{fig:CollapseInterfaceCondition1} and \ref{fig:CollapseInterfaceCondition2} are maintained. This is why these two rules need to be overridden and a new condition implemented: If two non-consecutive nodes (nodes not connected by the microstructural wireframe) collapse, the classification of the remaining node is a P-Node. Additionally, the remaining node is moved to the barycenter of the initial nodes involved in the collapse and the surrounding geometrical entities (points, lines and surfaces) are checked and updated if necessary. Take for example the same configuration shown in Fig. \ref{fig:CollapseInterfaceCondition1} now in Fig. \ref{fig:CollapseInterfaceCondition_Override_1}: here the collapsing of nodes $N_i$ and $N_k$ is possible, the remaining node $N_i$ is placed in the middle of the edge $\overline{N_i N_j}$ and its classification is changed from L-Node (blue) to P-Node (red). Now their surroundings need to be checked for possible changes on the topology: all three Lines (grain boundaries in green) need to erase node $N_k$ as a final/initial point and put in its place P-Node $N_i$, similarly, one of the lines has to add $N_k$ as a node in their sequence of L-Nodes hence $N_k$ changes also its classification to L-Node.

\begin{figure}[!h]
\centering
\includegraphics[width=1.0\textwidth] {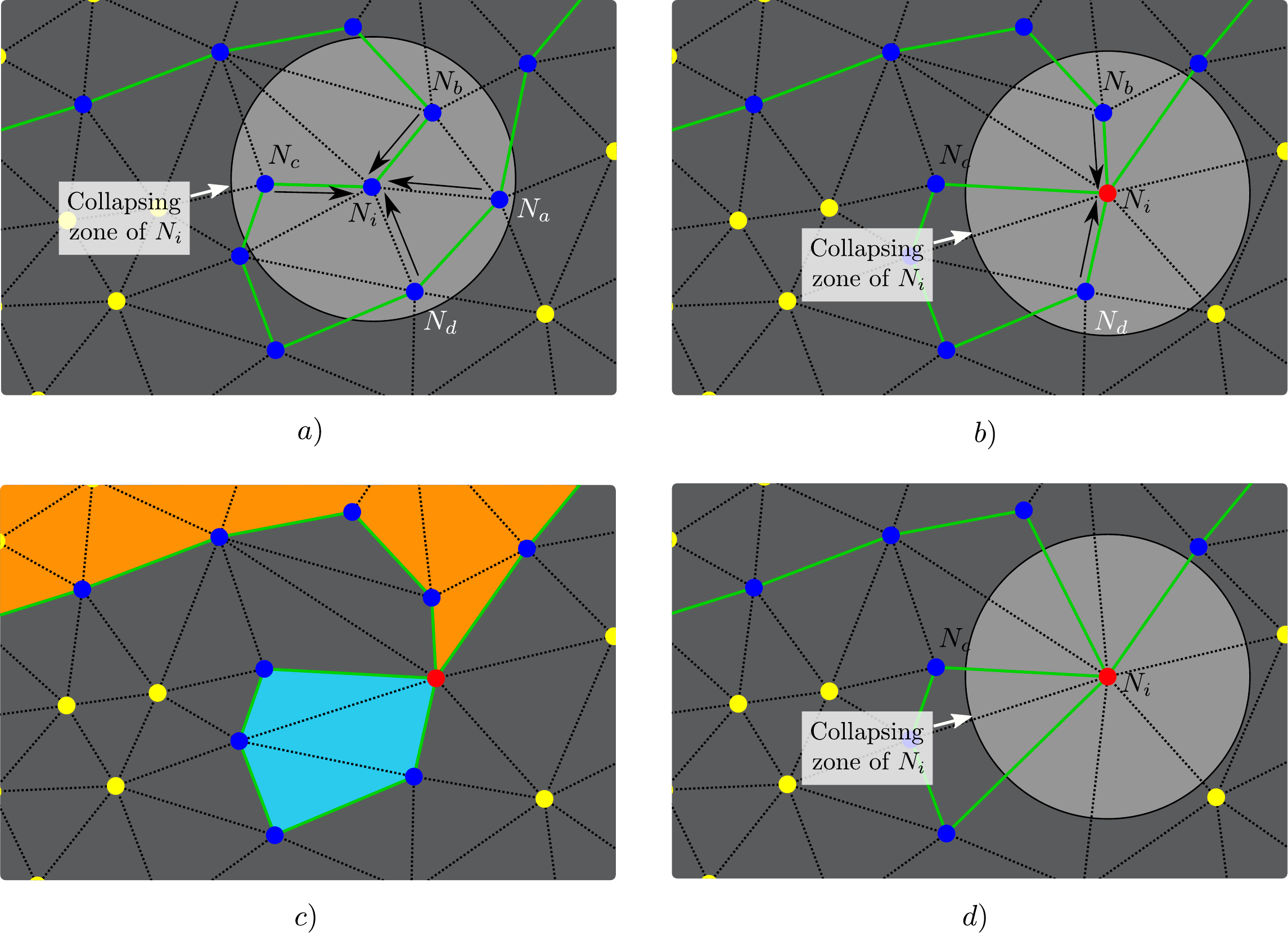}
\caption{Node Collapsing of Fig. \ref{fig:CollapseInterfaceCondition2} when allowing the collapse between non consecutive nodes, S-Nodes are displayed in yellow, L-Node in blue and P-Nodes in red. a) Initial state; b) L-Node $N_i$ performs the first collapse with $N_a$. This produces L-Node $N_i$ to be moved to the center of the edge $\overline{N_i N_a}$ and to become a P-Node. Moreover, the collapsing zone of $N_i$ changes its position and leave Node $N_c$ out; c) The collapse also has produced a Surface to be divided in two (cyan and orange Surfaces); d) Additional collapses are performed between P-Node $N_i$ and L-Nodes $N_b$ and $N_d$, these collapses are performed in the conventional way without moving $N_i$ as it is a P-Node now.}
\label{fig:CollapseInterfaceCondition_Override_2}
\end{figure}

Similarly, the situation presented in Fig. \ref{fig:CollapseInterfaceCondition2} can be reproduced with the new rules of collapsing: on the initial state (Fig. \ref{fig:CollapseInterfaceCondition_Override_2}a)), the collapsing zone of $N_i$ puts L-Nodes $N_a$, $N_b$, $N_c$ and $N_d$ inside. The first node to be collapse is L-Node $N_a$. After this initial collapse (Fig. \ref{fig:CollapseInterfaceCondition_Override_2}b)) the collapse produces L-Node $N_i$ to be moved to the center of the edge $\overline{N_i N_a}$ and to become a P-Node. Moreover, the collapsing zone of $N_i$ changes its position and leave L-Node $N_c$ out hence it will not be collapsed. Additionally, a new topological change is identified (Fig. \ref{fig:CollapseInterfaceCondition_Override_2} c)), here the collapse is also responsible of the fact that a Surface is divided in two new surfaces (cyan and orange Surfaces). Finally, the remaining collapses are performed between P-Node $N_i$ and L-Nodes $N_b$ and $N_d$ (Fig. \ref{fig:CollapseInterfaceCondition_Override_2}d)), these collapses are performed in the conventional way following the rules of collapsing between P-Nodes and L-Nodes presented in \cite{Florez2020b}.\\

The new node collapsing rules have been implemented in the TRM model and will be used from this point forward in the cases where SE is present. This node collapsing technique will be able to perform the majority of the topological changes in the microstructure. However not all topological changes can be handled by this node collapsing mechanism, indeed one special topological change needs to be handled differently: The creation of boundaries by the decomposition of unstable multiple junctions with more than 3 intersected boundaries. This phenomenon has been addressed in \cite{Florez2020b} section 3.2.4 where isotropic grain boundaries were considered, by the implementation of a new operator used on the mesh where such a configuration appears. This operator is responsible for the successive dissociation of grain boundaries from multiple junctions until a stable configuration is obtained. Here we will use the same algorithm denominated ``Split of multiple junctions".

\subsection{The TRM algorithm under the influence of capillarity and stored energy}

Finally the algorithm for a time step of the TRM model in the context of isotropic grain growth under the influence of stored energy and capillarity is presented in Algorithm \ref{alg:TRMGGPar}, where the step ``Perform Remeshing and Parallel Sequence" corresponds to the parallel implementation of the TRM model presented in \cite{Florez2020c}.

\begin{algorithm}
\caption{Isotropic Grain Growth TRM Algorithm for capillarity and stored energy}\label{alg:TRMGGPar}
\begin{algorithmic}[1]
\State \textbf{Perform Remeshing and Parallel Sequence} 

\ForAll{Points: $P_i$}
	\While{Number of Connections $>3$ }
		\State split multiple point $P_i$.
	\EndWhile
\EndFor
\ForAll{Lines : $L_i$}
    \State Compute the natural spline approximation of $L_i$.
\EndFor
\ForAll{L-Nodes : $LN_i$}
    \State Compute curvature and normal ($\kappa \vec{n}$) over $LN_i$ then compute $\vec{v_c}$ for $LN_i$ (Eq \ref{Eq:VelocityEquationc}).
    \State Compute the $\vec{v_e}$ for $LN_i$ (Eq \ref{Eq:VelocityEquatione})
\EndFor
\ForAll{P-Nodes : $PN_i$}
    \State Compute the product $\kappa \vec{n}$ over $PN_i$ using model II of \cite{kawasaki1989} then compute $\vec{v_c}$ for $PN_i$ (Eq \ref{Eq:VelocityEquationc}).
    \State Compute $\vec{v_e}$ for $PN_i$ (Eq. \ref{Eq:VelocityEquationeMP})
\EndFor

\State \textbf{Delete Temporal Nodes}

\ForAll{L-Nodes and P-Nodes : $LPN_i$}
	\State Compute final velocity $\vec{v}$ of Node $LPN_i$ (Eq. \ref{Eq:VelocityEquation})
    
\EndFor

\State \textbf{Iterative movement with flipping check in parallel} 
\end{algorithmic}
\end{algorithm}

\section{Recrystallization} \label{sec:ReX}

In order to model Recrystallization (ReX) with the TRM model, two additional components are necessary: the first is a procedure allowing to change the topology of the microstructure and to introduce new grains (i.e. nuclei); the second component is a model of apparition of nucleus which depends thermomechanical conditions. Here discontinuous dynamic ReX (DDRX) context is considered. Of course post-dynamic ReX (PDRX) and subsequent GG phenomena can also be investigated by considering microstructure evolutions when the deformation is completed. The combination of these two mechanisms can describe multiple TMTs that are used today in the material forming industry.

With the purpose of simplicity, in this article we will use the same methodology presented in \cite{Maire2017} for the laws governing the introduction of new nuclei during the modeling of hot deformation:

In \cite{Maire2017} the evolution of the dislocation density is accounted by a Yoshie-Laasraoui-Jonas Law \cite{Laasraoui1991} as follows: 

\begin{equation}
\label{Eq:EDPEvolutiondislocationdensity}
\centering
\dfrac{\partial \rho}{\partial \epsilon_{eff}^p} = K_1 - K_2 \rho, 
\end{equation} 

which can be evaluated in a discretized time space with an Euler explicit formulation for the next increment step as :

\begin{equation}
\label{Eq:DiscretEvolutiondislocationdensity}
\centering
\rho^{(t+\Delta t)} = K_1 \Delta \epsilon + (1 - K_2 \Delta \epsilon) \rho^{(t)},
\end{equation} 

where $\rho^{(t)}$ is the value of the dislocation density at time $t$ and where the value of $\Delta \epsilon$ can be computed as $\dot{\epsilon_{eff}^p} \cdot \Delta t$ with $\Delta t$ the time step.\\

As explained in \cite{Maire2017}, when a grain boundary migrates, the swept area is assumed almost free of dislocations. This aspect is modeled by attributing to these areas a value of dislocation density equal to $\rho_0$, then, for the grains with part of their domain presenting $\rho_0$, their dislocation density is homogenized within the grain (as intragranular gradients on the stored energy are not taken into account neither in \cite{Maire2017} nor in the present work), the final value of $\rho$ for the growing grains is computed as: 

\begin{equation}
\label{Eq:growingrainsdislocation}
\centering
\rho^{(t+\Delta t)} =\dfrac{ \rho^t S^(t) + \Delta S \rho_0}{S^{(t+\Delta t)}},
\end{equation} 

where $S^t$ and $\Delta S$ denote the surface at time $t$ and the change of surface of a given grain.\\

Additionally to Eq. \ref{Eq:growingrainsdislocation}, in PDRX the annihilation of dislocations by recovery must be taken into account. This is done thanks to the following evolution law: 

\begin{equation}
\label{Eq:annihilation}
\centering
\dfrac{d \rho}{d t}=-K_s \rho ,
\end{equation} 

where $K_s$ is a temperature dependent parameter representing the static recovery term. This recovery law is only taken into account in PDRX as in DRX, Eq. \ref{Eq:growingrainsdislocation} already takes into account the annihilation phenomenon.

\subsection{Nucleation laws} \label{sec:NucLaws}

The procedure consists of introducing volume (surface in 2D) of nuclei at a rate of $\dot{S}$, once the local value of dislocation density has reached a critical value: $\rho_c$. In \cite{Scholtes2016, Maire2017, Laura2014} this value was determinated by iterating until convergence the following equation:

\begin{equation}
\label{Eq:CriticRhoDRX}
\centering
\rho_c^{(i+1)^*} = \left [ \dfrac{-b \gamma \dot{\epsilon} \dfrac{K_2}{M \delta_{(\dot{\epsilon})} \tau^2 }}{ln(1-\dfrac{K_2}{K_1}{\rho_c^i})} \right ] ^{\dfrac{1}{2}}\text{ with } \rho_c^i = \rho_c^{i-1} + c \cdot (\rho_c^{(i)^*} - \rho_c^{i-1}),
\end{equation} 

where $i$ represents the iteration number, $c$ is a convergence factor ($c<1$ chosen in this article as $c=0.1$), $K_1$ and $K_2$ represents the strain hardening and the material recovery terms in the Yoshie–Laasraoui–Jonas equation discussed in \cite{Laasraoui1991}, the term $b=1$ in 2D and $b=2$ in 3D, $\tau$ is the dislocation line energy and $\dot{\epsilon}$ is the effective deformation rate used during the deformation of the material. \\

When solving this equation, two special cases may produce an erroneous computation: the first is given when $K_1/K_2*\rho_c > 1$ for which the logarithm is undefined, the solution to this is to limit the value of $\rho_c<K_2/K_1$ whenever this situation occurs. The second is when $\dot{\epsilon} = 0$ which correspond to the intervals where PDRX is considered. Two solutions may be considered for this situation: the first is to block the nucleation when it is not necessary (metadynamic evolution for example), and the second to supply value of $\dot{\epsilon}>0$ to Eq. \ref{Eq:CriticRhoDRX}. Here, we have chosen the latter, for which an apparent effective strain rate $\dot{\epsilon_s}$ is used instead $\dot{\epsilon}$ in PDRX: 

\begin{equation}
\label{Eq:StaticEpsilonEquation}
\centering
\dot{\epsilon_s} = \dfrac{\int_0^t{\dot{\epsilon}^2 ~ dt}}{\int_0^t{\dot{\epsilon} ~ dt}},
\end{equation} 

which accounts for the instant mean value of the real effective strain rate.\\

Once a value of $\rho_c$ is computed, the surface per unit of time $\dot{S}$ of nuclei to be inserted  can be computed with the following equation corresponding to a variant of the proportional nucleation model \cite{Peczak1993}:

\begin{equation}
\label{Eq:nucleationRateDrX}
\centering
\dot{S}= K_g P_c,
\end{equation} 

where the term $K_g$ is a probability constant depending on
the processing conditions and $P_c$ is the total perimeter of the grains whose dislocation density is greater than $\rho_c$.\\

Another constraint is given by the minimal radius $r^*$ of nucleation (the radius at which the nuclei should be inserted in the domain so the capillarity forces would not make it disappear) which can be computed thanks to the following equation \cite{Bailey1962}: 

\begin{equation}
\label{Eq:MinimalRadiusNuclei}
\centering
r^*=\omega\dfrac{\gamma}{(\rho_c -\rho_0)\tau},
\end{equation} 

where $\omega > 1$ is a safety factor ensuring the growing of the nucleus at the moment of its apparition. The term $\omega$ accounts for the non spherical shape of a grain inserted in a discretized domain such as in the TRM model. In section \ref{sec:metacircle} a value for this factor will be obtained based on numerical tests.

\subsection{Nucleation approach for the TRM model} \label{sec:Nucleation}

Till here we have defined the tools needed to obtain the kinetics of the grain boundaries, where the pressure behind such kinetics can be of different nature: capillarity, stored energy or both. However, in order to model ReX it is necessary to have a way to introduce new grains into the domain of the TRM model. Nucleation, similarly to boundary migration, is one of the ways of the microstructure to relax the high gradients of the stored energy appearing during or after a TMT. Nucleation has been addressed by several approaches for each methodology able to simulate such behavior: LS-FE methods, relay on the definition of circular LS fields (different from the already defined LS fields occupying the same spatial domain) to form nuclei \cite{Scholtes2016a, Maire2017}, CA and MC methods change the crystallographic orientation and SE value of some cells \cite{Hesselbarth1991, Davies1997, Sieradzki2013, Villaret2020} in order to nucleate and vertex models form new grains by redefining new vertex and interfaces in the shape of triangles around the pre-existent vertex points \cite{Mellbin2015}. \\

\begin{figure}[!h]
\centering
\includegraphics[width=0.9\textwidth] {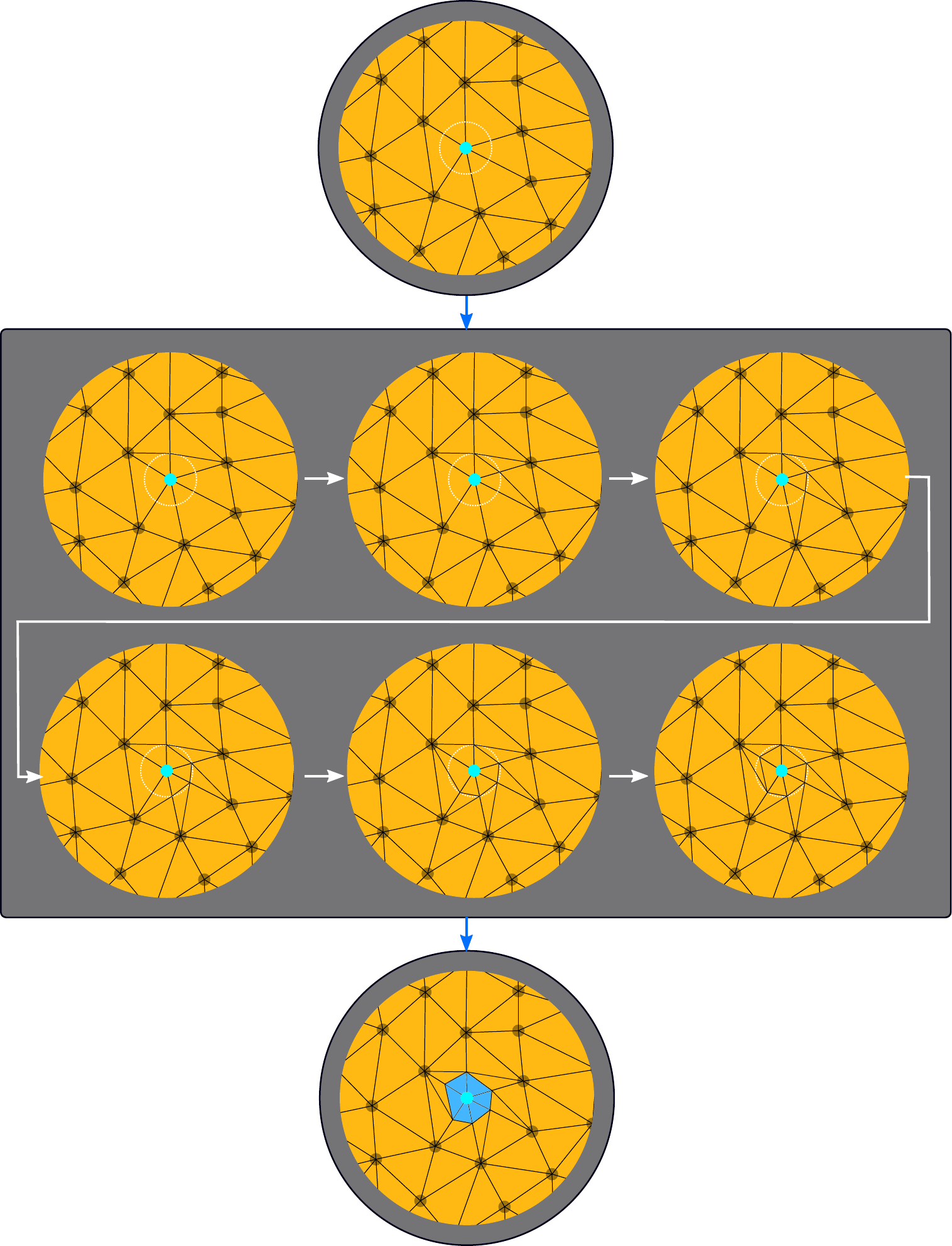}
\caption{Remeshing steps for the nucleation process of the TRM model. Top: initial state with a selected node (cyan) and a circle drawn over the mesh, middle: successive edge splitting steps to form the interfaces of the nucleus, bottom: the elements inside the nucleus are identified and extracted from its previous Surface container.}
\label{fig:Nucleation}
\end{figure}

\begin{figure}[!h]
\centering
\includegraphics[width=0.9\textwidth] {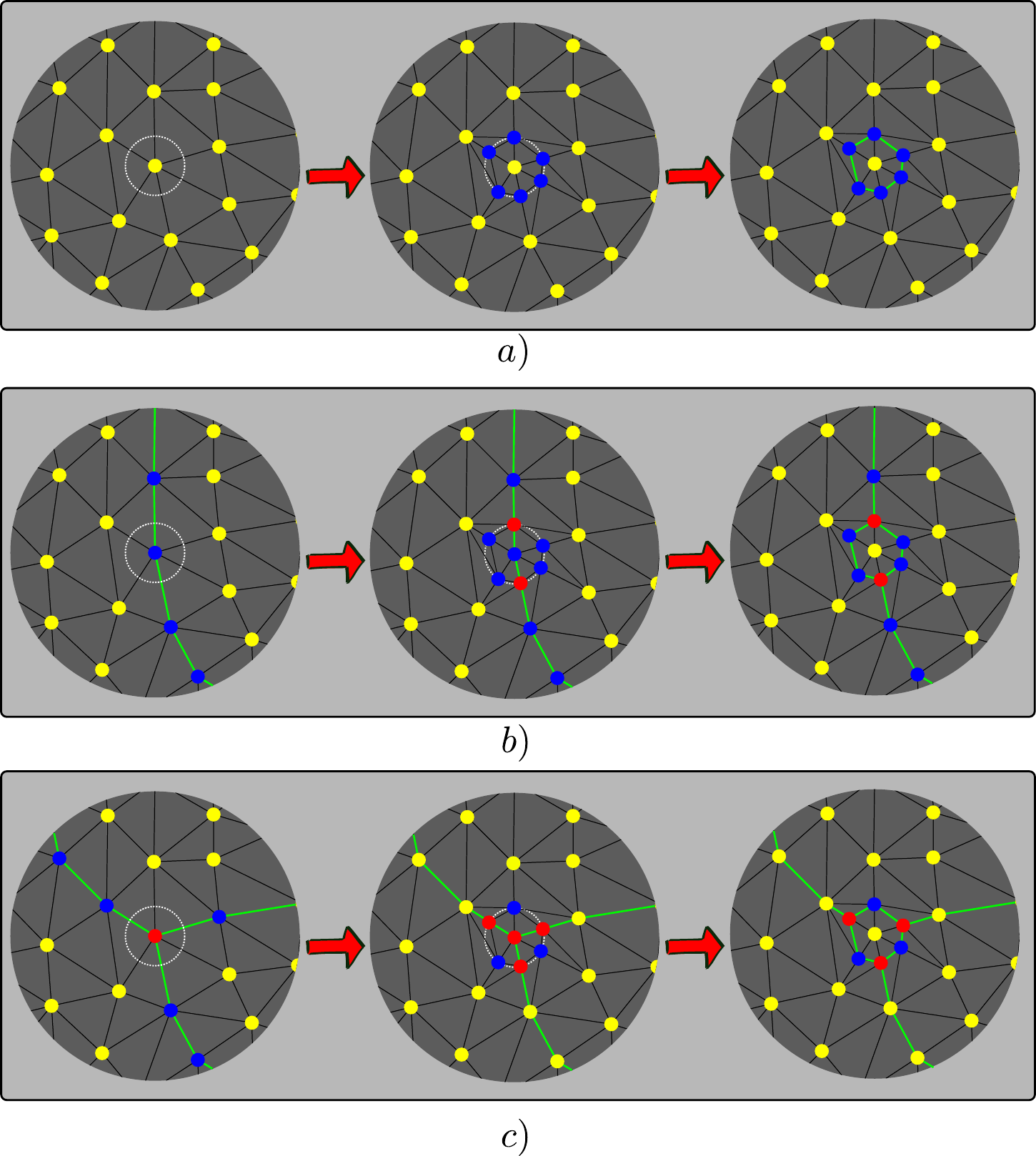}
\caption{Examples of the formation of nucleus over different types of nodes, a) S-Node at the center and no crossed lines, b) L-Node at the center and one crossed line, two P-Nodes are created, the initial line is divided and two new lines are created, c) P-Node at the center, 3 P-Nodes are created, the P-Node at the center is detached from all its lines and converted to S-Node, 3 new lines are created.}
\label{fig:NucleationPlaces}
\end{figure}

In the present work, a remeshing-reidentification procedure will be performed around a central node $N_i$ in order to introduce nuclei. A circular region with center $N_i$ will be drawn and all edges crossed by this circle will be split at the intersection in a similar manner as in \cite{Shakoor2017ijnme, Florez2020} by successively applying an edge splitting operation, regardless of the classification of the nodes defining the edge (P-Node/L-Node/S-Node) (see Fig. \ref{fig:Nucleation}). The classification of the new nodes being placed by the splitting algorithm are as L-Nodes unless the split edge represents a grain boundary, in which case the inserted node will be classified as a P-Node  (see Fig. \ref{fig:NucleationPlaces} b) middle and c) middle). Once all edges are split, a Surface Identification algorithm will be performed over node $N_i$ (see section 2.2.5 of \cite{Florez2020b}), all identified elements and nodes will be inserted into a new empty Surface defining the nucleus, and extracted from their previous Surfaces (grains), new Lines (grain boundaries) will be built with their respective Points (multiple junctions) if any were formed by the nucleation process and all remaining lines and points inside the new surface will be destroyed (remaining lines and points can appear if the nucleation took place near over one grain boundary) see Fig. \ref{fig:NucleationPlaces} a), b) and c) left. However in a parallel context, an additional constraint was added: shared nodes can not be involved in the nucleation process, neither as a central node nor one of the nodes of a split edge. This constraint was added because of the lack of information (position of the edges to cut) around shared nodes and the performance of the nucleation process (as a great amount of information would be necessary to be transferred to other processors). This constraint should not have a great impact on the general behavior of the model as the domain of each processor (and their shared nodes) is changed by the \emph{Unidirectional Element Sending} algorithm presented in \cite{Florez2020c} every time step, hence constantly unblocking the restriction to nucleate over the same region.

%% file: NumericalResults.tex
\section{Numerical Tests}

In this section different academic tests will be performed to evaluate the performance of the TRM model when simulating GBM under the influence of capillarity and stored energy. For these academic tests, adimensional simulations will be considered. Moreover, results of  simulations using the DRX and PDRX frameworks described in section \ref{sec:ReX} will be given, these simulations will use the nucleation approach presented in section \ref{sec:Nucleation} specially developed for the TRM model. The different physical parameters will be take as representative of the 304L stainless steel. Comparisons with LS-FE predictions will be discussed.

\subsection{Circular Grain: competition between capillarity and stored energy}\label{sec:metacircle}

\begin{figure}[!h]
\centering
\includegraphics[width=0.8\textwidth] {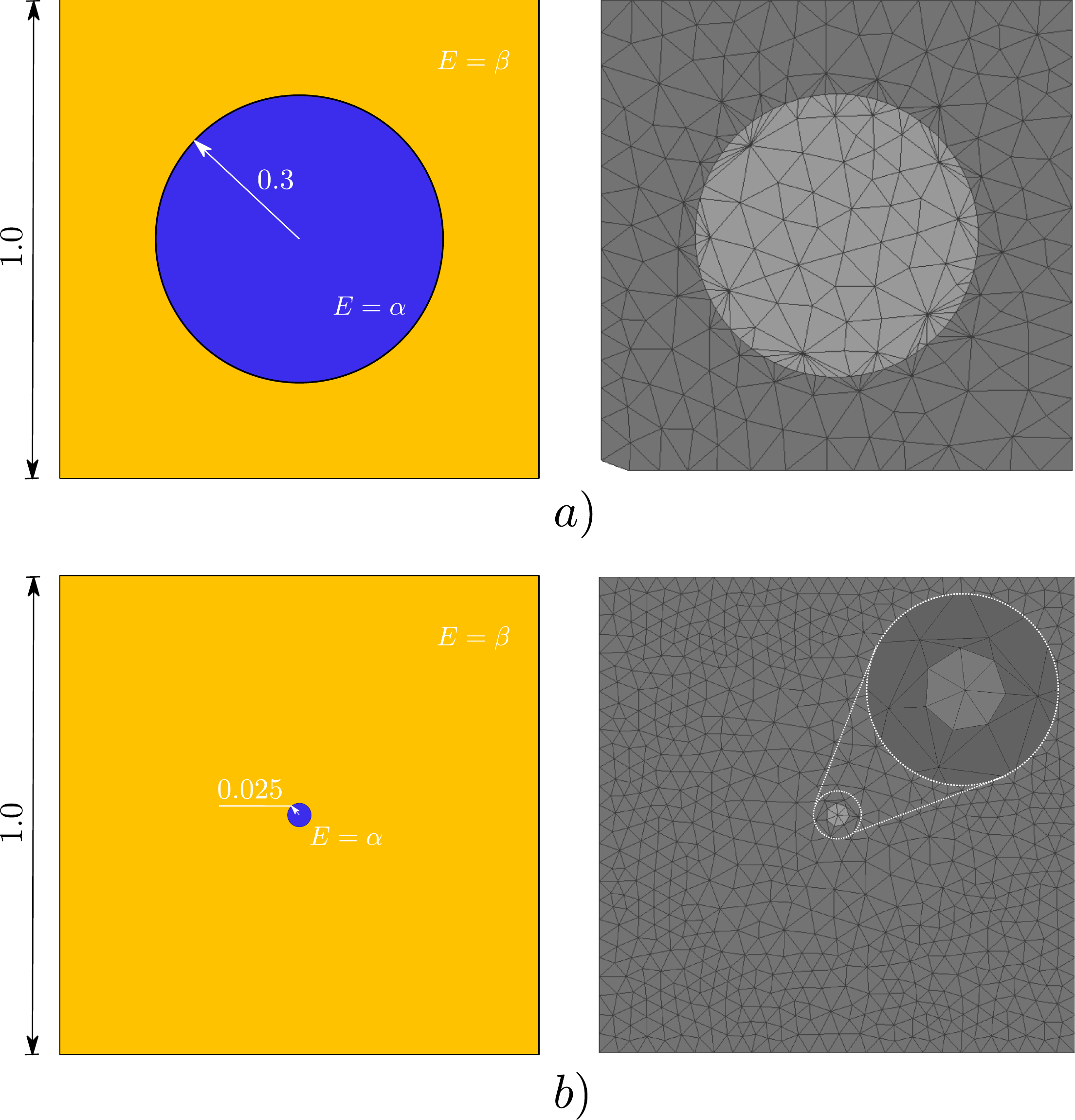}
\caption{Circle Test, left: initial state and right: initial mesh a) $r=0.3$ radius (Surface=$0.287$) b) $r=0.025$ radius (Surface=$0.01963$)}
\label{fig:Circle_InitialState}
\end{figure}

In this test case, it will be evaluated the accuracy of the model when the geometric configuration leads to a competition between the driving forces given by the capillarity and the SE. Here we will adopt a value of boundary energy and mobility equal to $\gamma=1$ and $M=1$ respectively. A circular domain with a value of stored energy $E=\alpha$ is immersed in a squared domain with an attributed value of stored energy $E=\beta$ ( see Figure \ref{fig:Circle_InitialState} a) and b) left) where $\beta>\alpha$. The difference on the stored energy $[E] = \beta-\alpha$  at the boundary will try to make the circle expand at a rate $v_e=M [E]=[E]$ while the capillarity effect will try to make it shrink at a rate $v_c=M\kappa=\kappa$ where $\kappa$ is the local curvature. The analytical model for this configuration can be put in terms of a non-linear ordinary differential equation in terms of the radius $r$ of the circle as follows: 

\begin{equation}
\label{Eq:nlodeRadiusCircle}
\centering
\dfrac{dr}{dt} = -\dfrac{1}{r} + [E],
\end{equation}

or in terms of the surface $S$ of the circle:

\begin{equation}
\label{Eq:nlodeSurfaceCircle}
\centering
\dfrac{dS}{dt} =  2 (- \pi + \sqrt{\pi S} [E]),
\end{equation}

We have used an Euler explicit approach to solve this equation and the results are used to compare the response of the TRM model for different values of $[E]$ for two cases: the first is given for an initial radius of $r_0=0.3$ and the second for $r_0=0.025$ (see Figure \ref{fig:Circle_InitialState} a).left and b).left). The initial mesh for each one of the two cases is given in Figure \ref{fig:Circle_InitialState} a).right and b).right respectively. Note how in the first case, the initial circle boundary is discretized by an amount of nodes sufficiently capable of capturing precisely the value of its curvature, hence it will serve to evaluate the accuracy on the kinetics of a typical curved boundary, while in the second case, the circle boundary is only defined by a few nodes allowing to evaluate the behavior of a nucleus when it is inserted on the domain.\\

\begin{figure}[!h]
\centering
\includegraphics[width=1.0\textwidth] {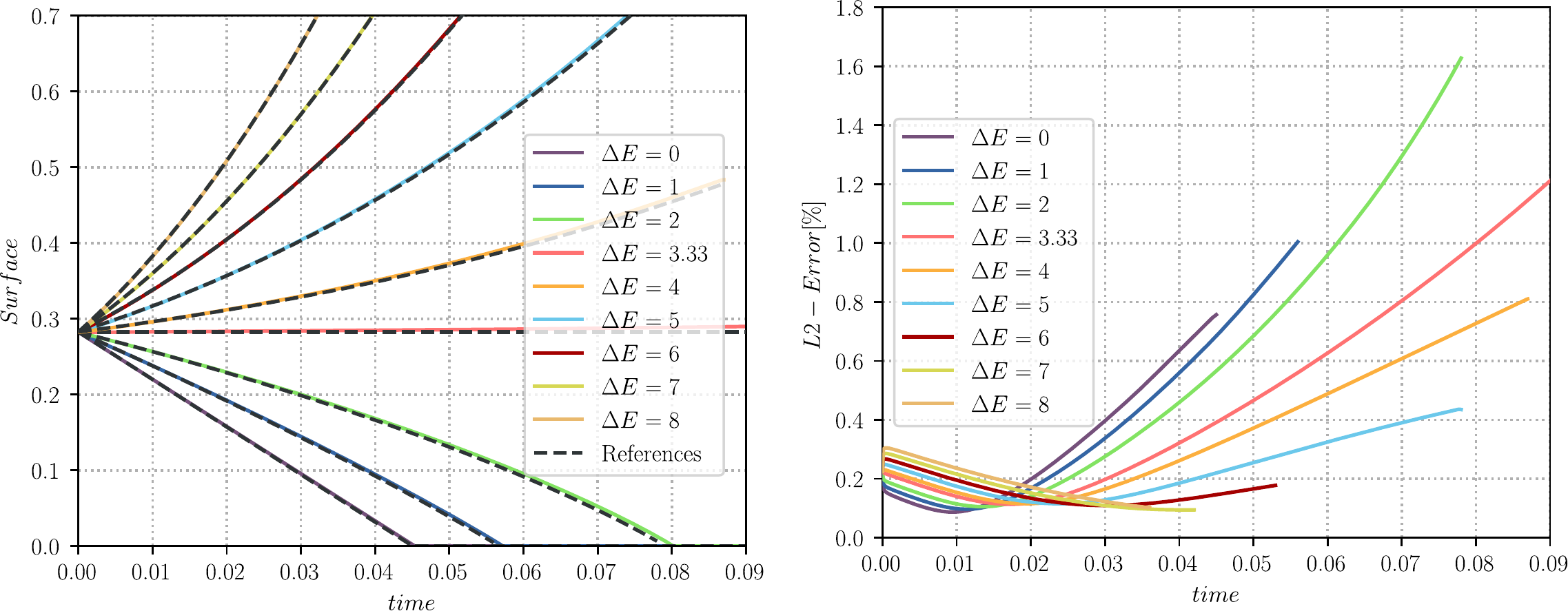}
\caption{Evolution of the surface (left) and L2 error (right) for the circle test case for an initial circle radius $r=0.3$ (Surface=$0.287$) a mesh size $h=0.006$ and a delta time $dt=3e-5$, the analytical results (References) are shown superposed to the simulated curves in black dashed lines. The expected metastable curve is given for a $[E] = 10/3 $ (Red curve). }
\label{fig:All_Sphere_Big}
\end{figure}

Results for this first case are given in Fig.~\ref{fig:All_Sphere_Big} along with the solution of Eq.~\ref{Eq:nlodeSurfaceCircle} for different values of $[E]$, Fig.~\ref{fig:All_Sphere_Big}.left illustrates how the references curves are superposed to the different simulated curves with a very low error (around 2 \% max see Fig. \ref{fig:All_Sphere_Big}.right). Furthermore, the analytic metastable case (given for $[E] = 10/3 $) shows a very good behavior losing only $1.2\%$ of its surface at $t=0.09$.

\begin{figure}[!h]
\centering
\includegraphics[width=1.0\textwidth] {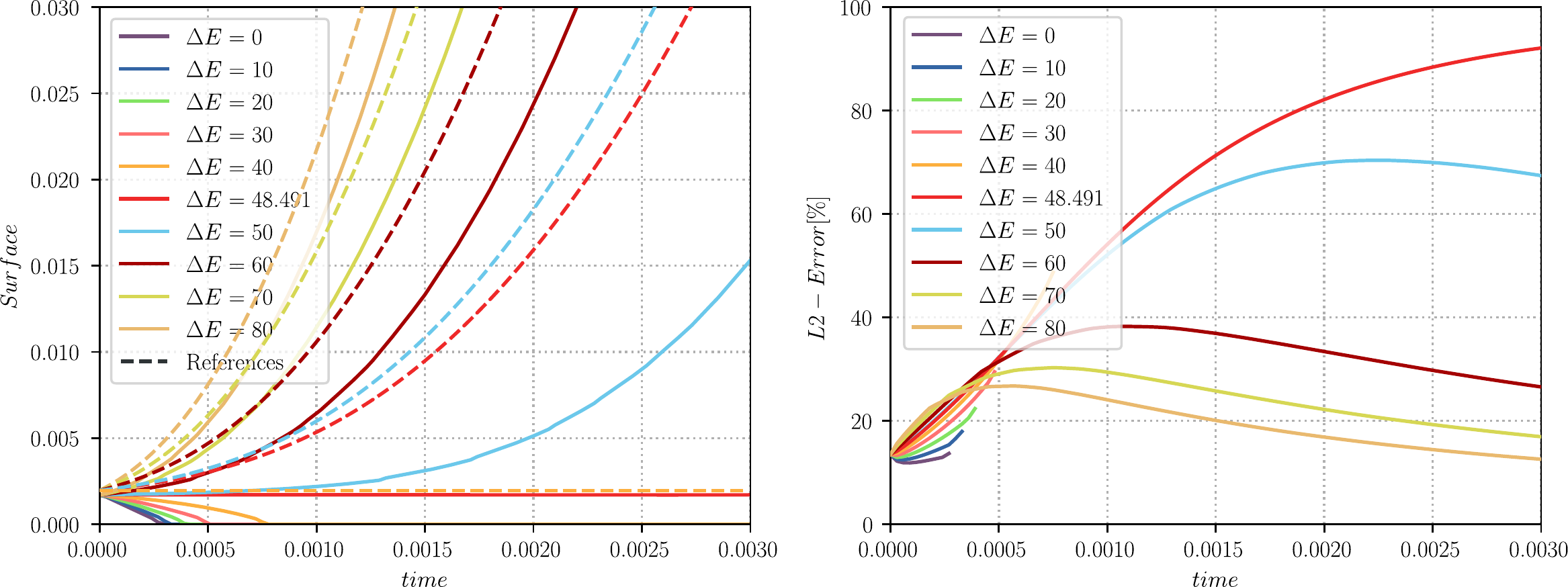}
\caption{Evolution of the surface (left) and L2 error (right) for the circle test case with an initial circle radius of $r=0.025$ (Surface=$0.01963$) a mesh size $h=0.025$ and a delta time $dt=1e-5$, the analytical results (References) are shown as dashed lines of the same color of their corresponding simulated evolution. The expected metastable curve is given for a $[E] = 40 $ (orange curve), but metastability was found for $[E] = 48.491$ (Red curve) }
\label{fig:All_Sphere_Small}
\end{figure}

Similarly, The results for the second circle case are given in figure \ref{fig:All_Sphere_Small}. Here it is appreciated how for the cases where the capillarity is the higher driven force ($[E]= 0, 10, 20, 30$), the circle disappears at the good rate. Interesting discussion concerns the case with $[E]= 40$ which corresponds analytically to the metastable configuration. In TRM simulation, the grain disappears. This behavior is due to the low number of nodes at the interface, producing an overestimation on the computed value of its curvature, making it shrink from the very first increment. A value of $[E]\approx 48,5$ was necessary on the simulated side to maintain a metastable position (an increase of $21.2\%$ accordingly to its analytical value). Moreover, for this value, the error on the prediction of the evolution of the surface was also the highest, going up to $92\%$ after $t=0.003$. Of course, this error is given as the simulated circle maintains its surface, while the analytical solution shows a continuous increase. The curves corresponding to $[E]= 50, 60, 70, 80$ (for which the higher driving force is the stored energy) show a decreasing error when the value of $[E]$ increases. This result can be used on the determination of factor $\omega$ used in Equation \ref{Eq:MinimalRadiusNuclei} where the authors have estimate that a value of $\omega=1.5$ (which counteracts for an increase of $50\%$ over the analytical value of $[E]$ for a metastable state, see the curve $[E] = 60$ in figure \ref{fig:All_Sphere_Small}.) is sufficient in order to give the inserted nuclei a growing state and prevent its early disappearance.

\subsection{Triple junction : The capillarity effect on the quasi-stable shape of multiple junctions }

In \cite{Reitich1996, Taylor1995} analytic solutions for the movement of multiple junctions in a quasi steady-state under the influence of stored energy were presented. In \cite{Reitich1996} the so called "Vanishing Surface Tension" (VST) test was introduced to demonstrate the non-uniqueness of the solution presented in \cite{Taylor1995} hereafter called the "Sharp" solution, this test (the VST test) takes the form of the limit problem given by:

\begin{equation}
\label{Eq:VelocityEquationVST}
\centering
\vec{v} \cdot \vec{n}=-M ( [E]_{ij} +\epsilon \gamma\kappa), ~\text{with}~ \epsilon \rightarrow 0,
\end{equation}

which has subjected to several 2D test cases and a perturbation analysis to demonstrate that the VST solution correspond to one of the solutions when $\epsilon=0$ and to the unique solution otherwise.\\

\begin{figure}[!h]
\centering
\includegraphics[width=0.95\textwidth] {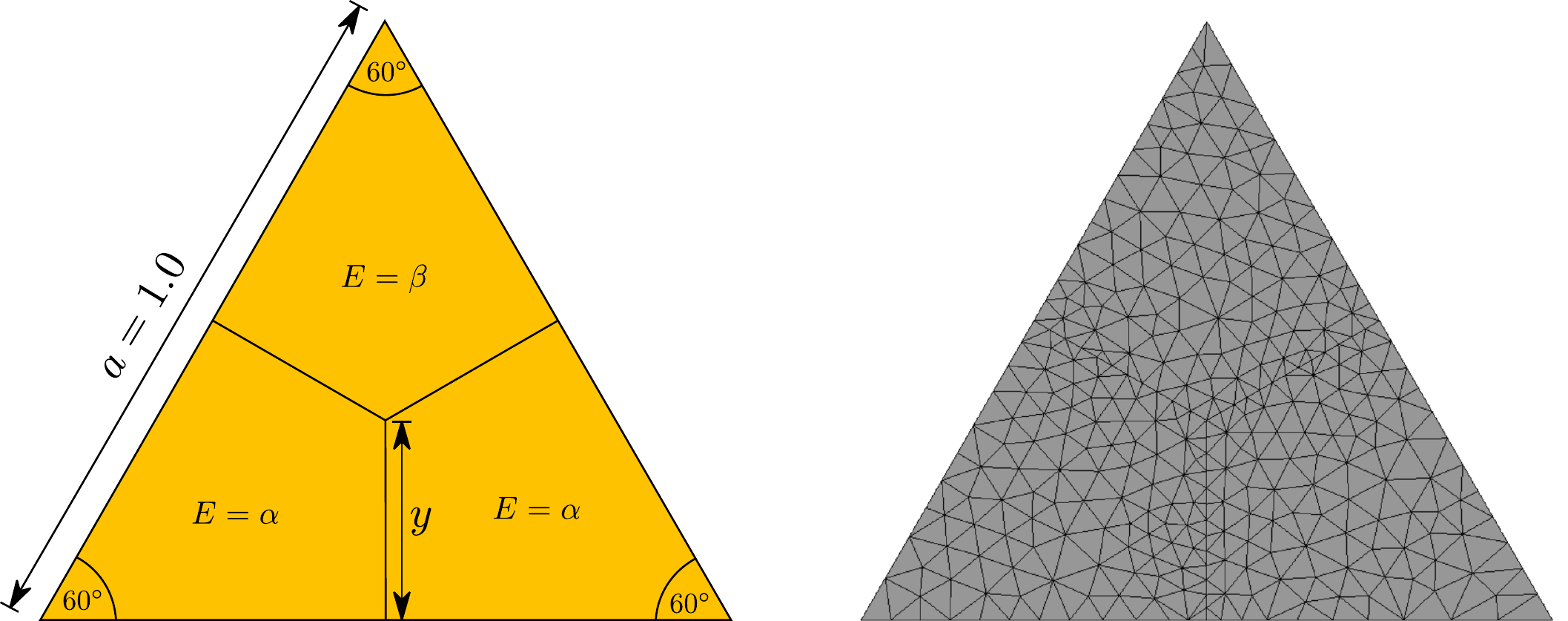}
\caption{Initial state for the triple junction test, three phases immersed in a domain in the shape of an equilateral triangle, this shape is intended to maintain an orthogonal position of the boundaries with respect of its limits while the configuration evolves. a) Initial configuration and b) initial mesh.}
\label{fig:MultJunc_InitialState}
\end{figure}

These solutions were later studied in \cite{Bernacki2008, Bernacki2009} using a LS-FE model to obtain the same behavior both in 2D and 3D. Here we have reproduced with the TRM model two tests that show the same behavior as in \cite{Reitich1996, Bernacki2008, Bernacki2009} for the 2D solutions. For all test the "Sharp" solution was obtained when capillarity effects where taken into account (with $\epsilon = 1$) and the VST solution when no capillarity was introduced in the system (hence with a value of $\epsilon=0$). Furthermore we have developed analytic equations for the evolution of the growing surface in our specific case (see Figure \ref{fig:MultJunc_InitialState}), these analytic evolutions are valid up to the point of contact of the multiple junction with the lower edge of the equilateral triangle (the limits of out domain) and allow us to make a more quantitative comparison in terms of error.\\

For this test case, the initial conditions are those presented in figure \ref{fig:MultJunc_InitialState}.left, three phases immersed in a domain in the shape of an equilateral triangle, this shape is intended to maintain an orthogonal position of the boundaries with respect of its limits while the configuration evolves. Two of the phases (the two in the lower part of the domain) will have a constant value of stored energy of $\alpha$ and the third phase a value of $\beta<\alpha$, this configuration will produce a global movement of the triple junction downwards at a constant and normal velocity of the flat interfaces equals to $\alpha-\beta$. Eventually the triple point will reach the bottom part of the domain making it to split and evolve towards a lower energy state; even though this portion of the simulation is showed in some of the results it is not relevant to our study, hence we will give quantitative results up to the point of splitting. The initial mesh for every test performed is shown in figure \ref{fig:MultJunc_InitialState}.right corresponding to a mesh size parameter of $h=0.006$. Furthermore, values for the boundary energy and mobility have been set to $\gamma=1$ and $M=1$ respectively.\\

The analytic solution for the evolution of the surface of the upper phase (the growing phase) for the Sharp solution is given by: 

\begin{equation}
\label{Eq:SurfaceTriplePointAnalytic_Cap}
\centering
S_{Cap} = \left(\dfrac{2a}{\sqrt{3}}-y \right)^2\dfrac{\sqrt{3}}{4}
\end{equation}

where $a$ is the length of one of the sides of the equilateral triangle (here $a=1$) and $y$ is the vertical position of the triple junction measured from the base of the triangle and given by the following expression: 

\begin{equation}
\label{Eq:yposition_triplepoint}
\centering
y = \dfrac{a}{2\sqrt{3}} - |\vec{v} \cdot \vec{n}| \dfrac{2t}{\sqrt{3}}
\end{equation}

where $t$ is the time and the expression $|\vec{v} \cdot \vec{n}|$ is the instant normal velocity of the flat phase boundaries, i.e. $\alpha - \beta$.\\

Similarly the analytic response of the VST solution in terms of surface for the growing phase is given by 

\begin{equation}
\label{Eq:SurfaceTriplePointAnalytic_NoCap}
\centering
S_{NoCap} = S_{Cap} + \left( \dfrac{\pi}{6} - \dfrac{1}{\sqrt{3}}\right)(|\vec{v} \cdot \vec{n}| t)^2.
\end{equation}

\begin{figure}[!h]
\centering
\includegraphics[width=0.6\textwidth] {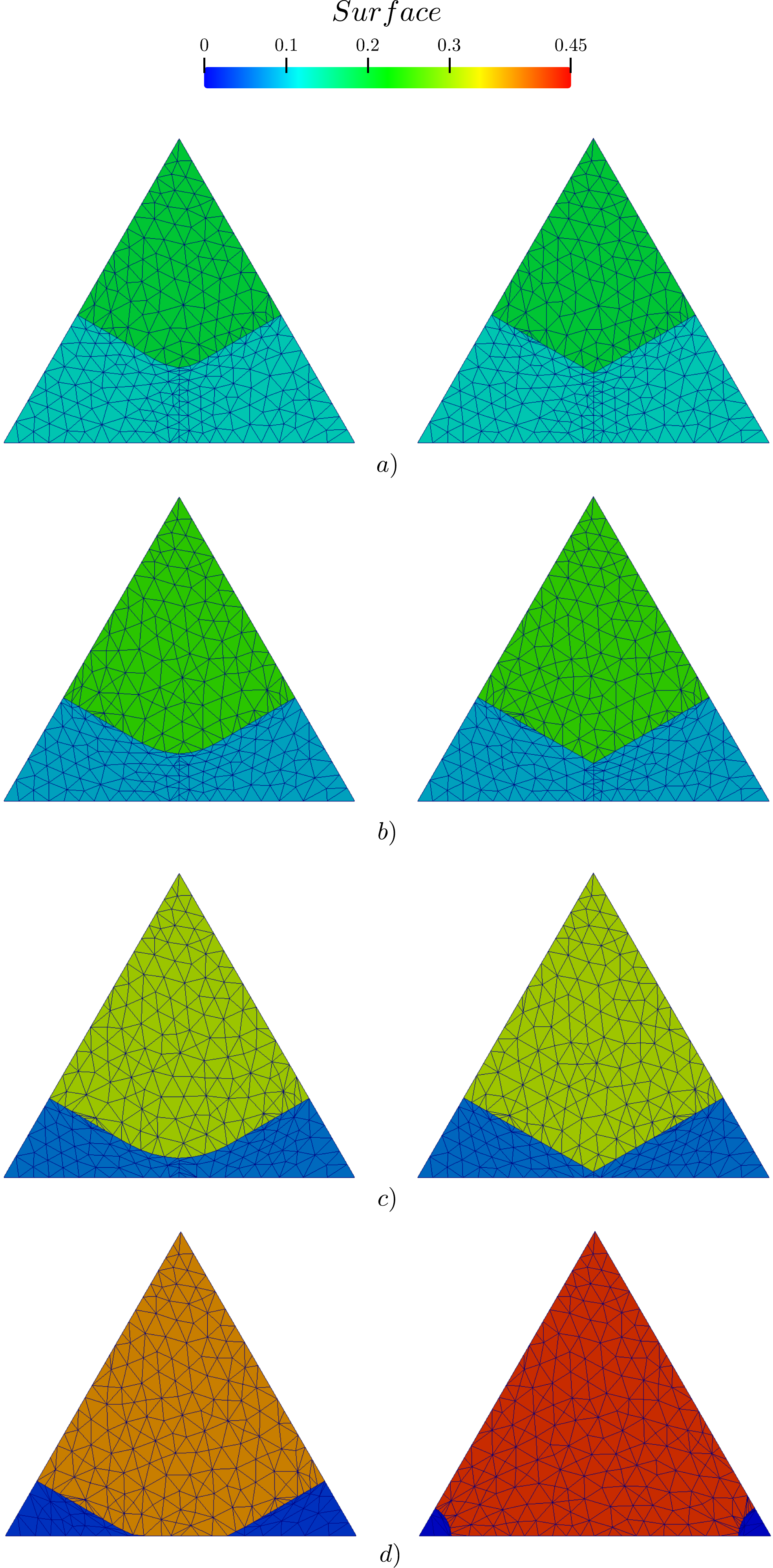}
\caption{States for the triple junction test case with a value of $[E] = 2$, left: with $\epsilon=0$ and right: with $\epsilon=1$ at a) $t=0.02$ b) $t=0.04$, c) $t=0.06$, d) $t=0.08$.}
\label{fig:MultJunc21}
\end{figure}

\begin{figure}[!h]
\centering
\includegraphics[width=0.6\textwidth] {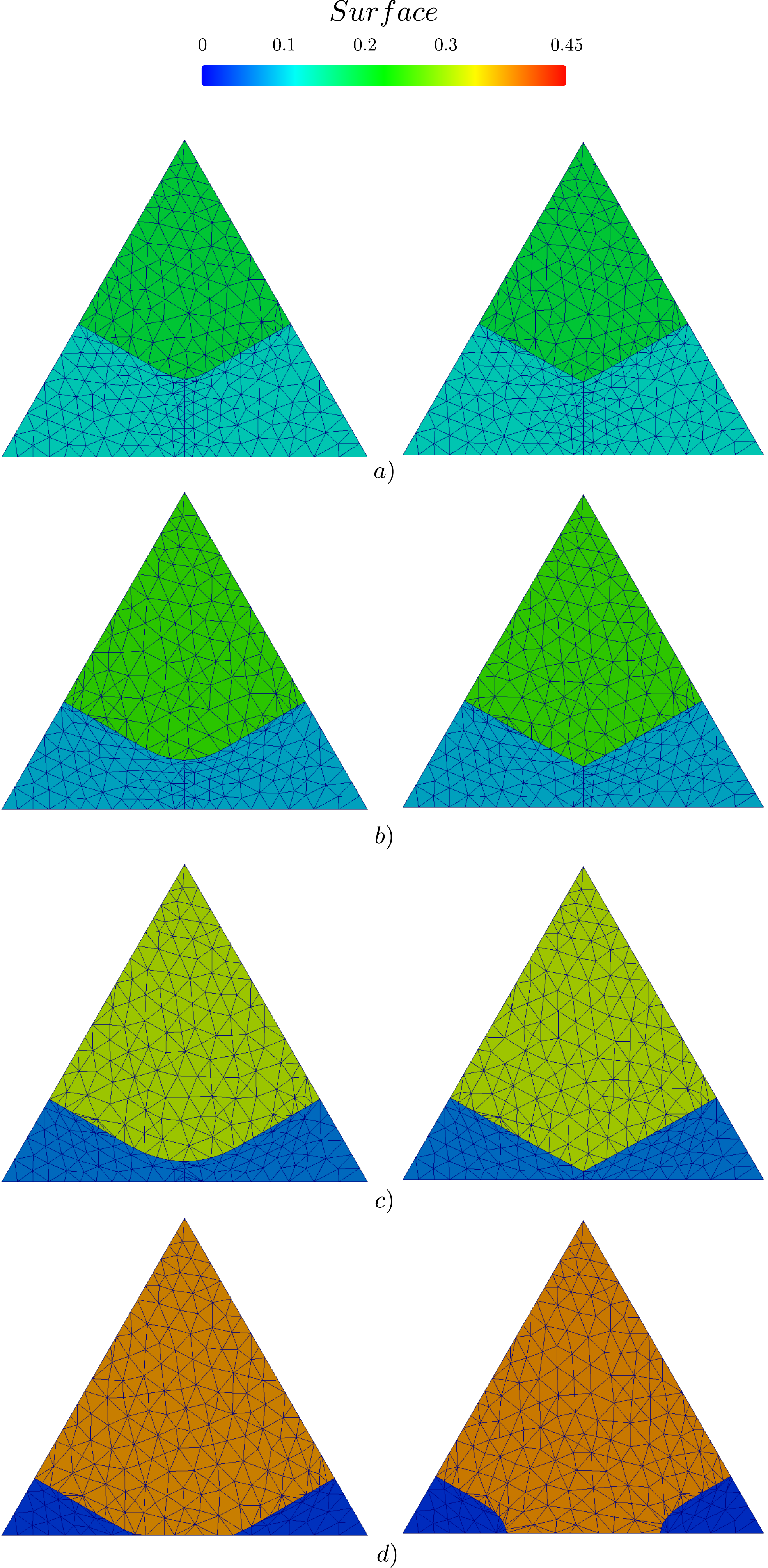}
\caption{States for the triple junction test case with a value of $[E] = 10$, left: with $\epsilon=0$ and right: with $\epsilon=1$ at the instant a) $t=0.004$ b) $t=0.008$, c) $t=0.012$, d) $t=0.016$.}
\label{fig:MultJunc105}
\end{figure}

Two test were performed: one with $\beta=2$ and $\alpha=4$, i.e. $|\vec{v} \cdot \vec{n}| = [E] = 2$ and one with $\beta=10$ and $\alpha=20$, i.e. $|\vec{v} \cdot \vec{n}| = [E] = 10$. The two tests were performed with a time step $\Delta t = 1 \cdot 10^{-5}$. Results for the evolution of the mesh and the surface are given in figures \ref{fig:MultJunc21} and \ref{fig:MultJunc105} for the first and the second case respectively. It is clear that the accuracy on the scalability of the solution is very good as figures \ref{fig:MultJunc21} a), b) and c) are almost equal to the ones of figures \ref{fig:MultJunc105} a), b) and c) respectively which were obtained for a velocity 5 times higher. Note that the only different frame is given for Figures \ref{fig:MultJunc21} d).right and \ref{fig:MultJunc105} d).right as here the capillarity effects over the limits of the domain are not negligible and in Figure \ref{fig:MultJunc21} d).right the configuration have had 5 times more time to evolve to its given state.\\

\begin{figure}[!h]
\centering
\includegraphics[width=0.9\textwidth] {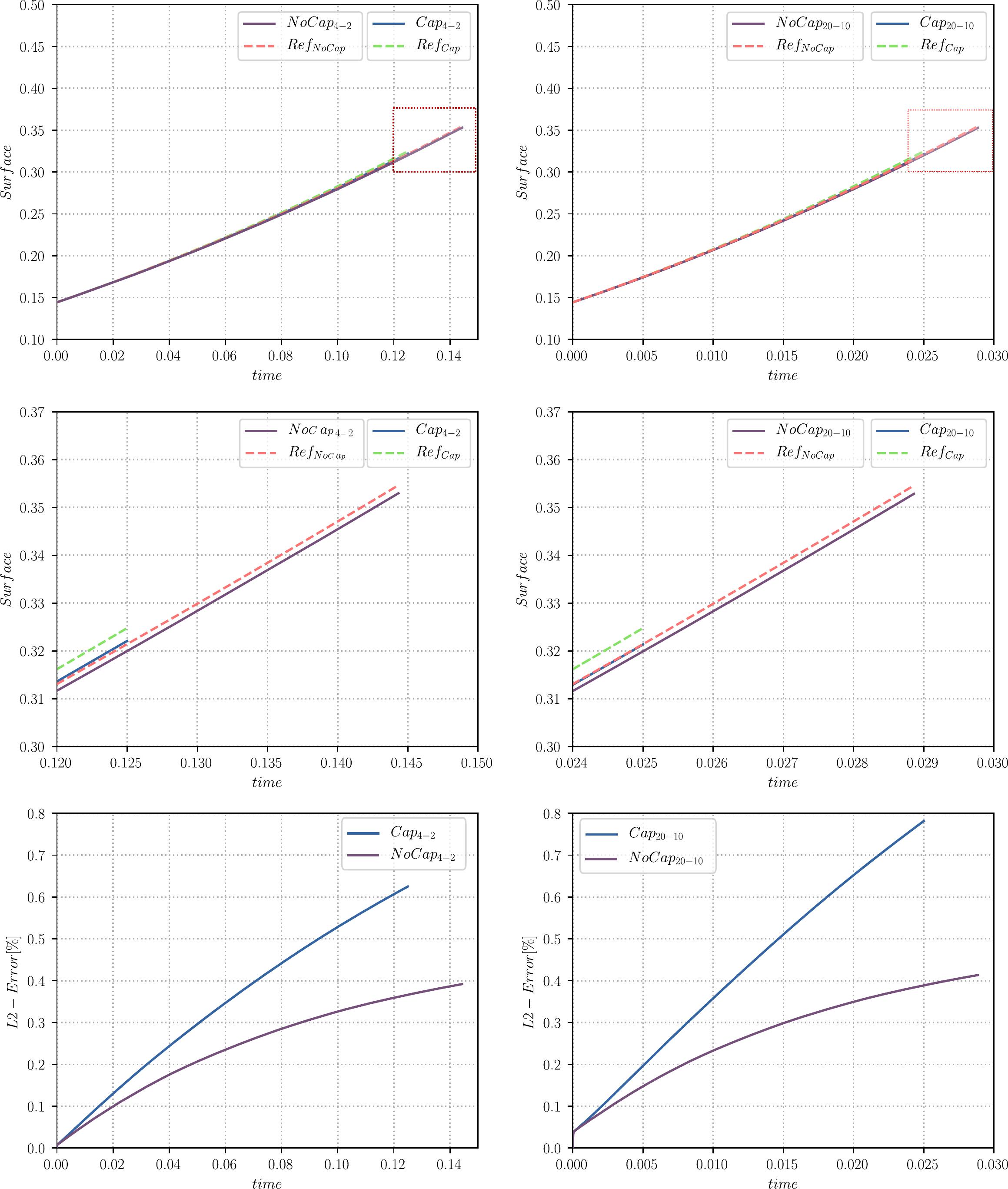}
\caption{Evolution of the surface of the growing phase of the triple junction test case, from top to bottom: (top) Evolution of the surface, (center) Zoom in the red zone, (bottom) L2-error over the evolution of the surface. Left: results for the test with $[E] = 2$ and right: with $[E] = 10$ }
\label{fig:ChartMultJunc_20}
\end{figure}

The evolution of the surface of the growing phase and its error with respect to equations \ref{Eq:SurfaceTriplePointAnalytic_Cap} and \ref{Eq:SurfaceTriplePointAnalytic_NoCap} is given in Figure \ref{fig:ChartMultJunc_20}, where the L2-error for both cases was lower than $0.8\%$.

\subsection{DRX/PDRX case}

\begin{figure}[!h]
\centering
\includegraphics[width=0.6\textwidth] {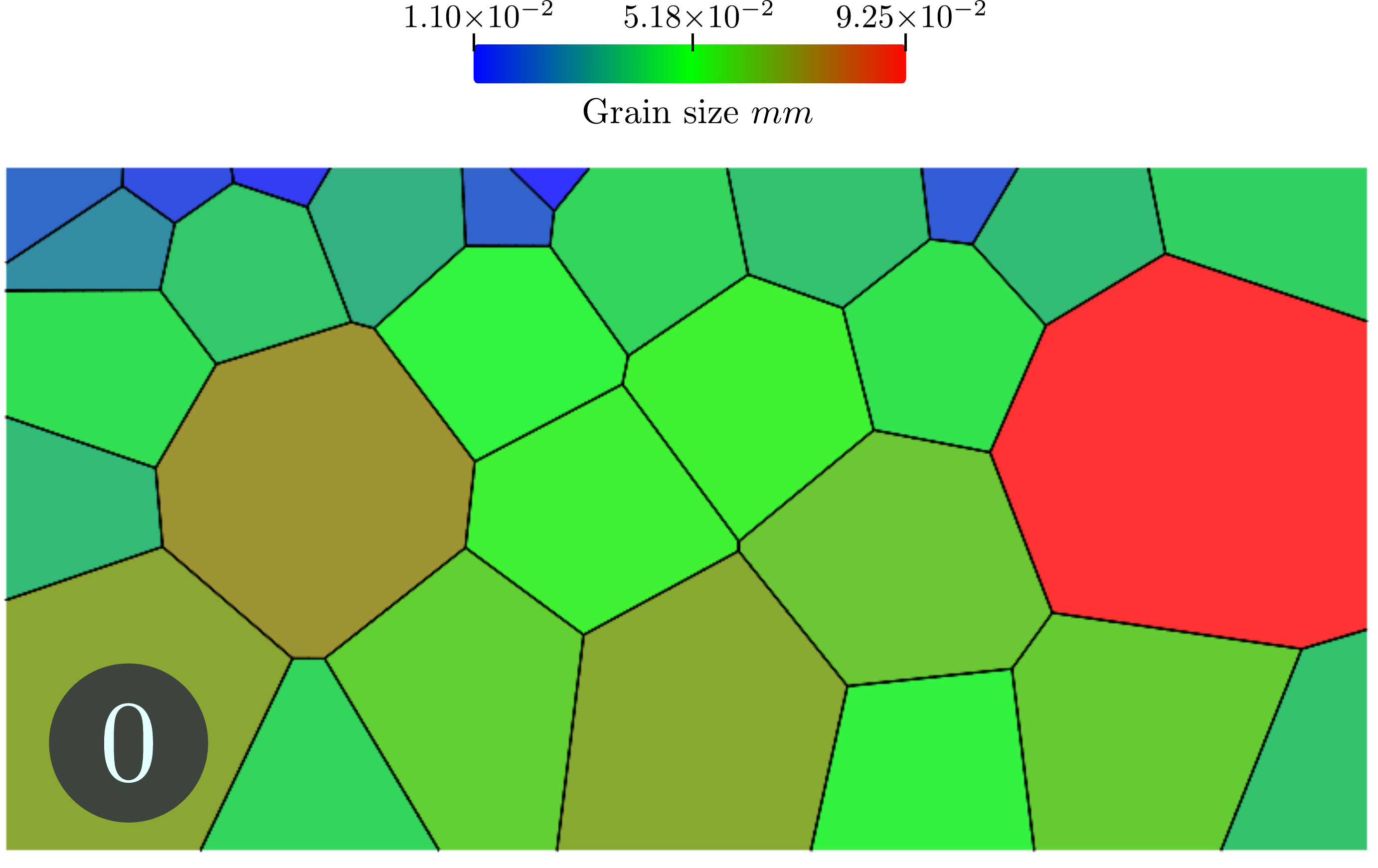}
\caption{Initial State for the DRX/PDRX test case.}
\label{fig:InitialStateDrX}
\end{figure}

Here a simulation with a few initial grains will be performed using the recrystallization method mentioned in section \ref{sec:ReX}: the initial tessellation will be realized thanks to a Laguerre-Voronoi cells generation procedure \cite{imai1985voronoi,Hitti2012,Ilin2016} over a rectangular domain of initial dimensions $0.65\times 0.328~mm$ (see figure \ref{fig:InitialStateDrX}) and the values for $M$, $\gamma$, $\tau$ and $k_s$ are chosen as representative of a 304L stainless steel at $1100$ $^{\circ}C$  (with $M=M_0*e^{-Q/RT}$ where $M_0$ is a constant $M_0=1.56\cdot 10^{11}$ $mm^4/Js$, $Q$ is the thermal activation energy $Q=2.8\cdot 10^5$ $J/mol$, $R$ is the ideal gas constant, $T$ is the absolute temperature $T=1353$ $K$, $\gamma=6\cdot 10^{-7}$ $J/mm^2$, $\tau=1.28331\cdot 10^{-12}~J/mm$ and $k_s= 0.0031~s^{-1}$ \cite{Scholtes2016, Maire2017} ). Additionally, the parameters $K_1$, $K_2$, $K_g$ and $\delta$ are taken as dependent of the absolute value of the component xx of the strain rate tensor $\dot{\varepsilon}$ ($|\dot{\varepsilon}_{xx}|$) which is defined as corresponding to a plane deformation case. These parameters will be obtained using a linear interpolation of the values presented in Tab. \ref{table:DataTable}.\\

\begin{table}[h]

\caption{Parameter data table for the DRX PDRX test case, when in range $|\dot{\varepsilon}_{xx}|=[0.01, 0.1]~s^{-1}$ the values are interpolated. If $|\dot{\varepsilon}_{xx}|>0.1$ the value for the corresponding parameter will the same as for $|\dot{\varepsilon}_{xx}|=0.1~s^{-1}$, the same strategy applies when $|\dot{\varepsilon}_{xx}|<0.01$. \label{table:DataTable}}

\begin{center}
\begin{tabular}{ |c|c|c|c|c| } 
 \hline
$|\dot{\varepsilon}_{xx}|~s^{-1}$ & $K_1~mm^{-2}$ & $K_2$ & $K_g~mm\cdot s^{-1}$ & $\delta$\\
 \hline
0.01 & 1.105 $\cdot 10^9$ & 9   & 1.3 $\cdot 10^{-4}$ & 0.937\\
0.1  & 1.55  $\cdot 10^9$  & 6.9 & 9  $\cdot 10^{-4}$ & 2.245\\

\hline
\end{tabular}
\end{center}

\end{table}

\begin{figure}[!h]
\centering
\includegraphics[width=1.0\textwidth] {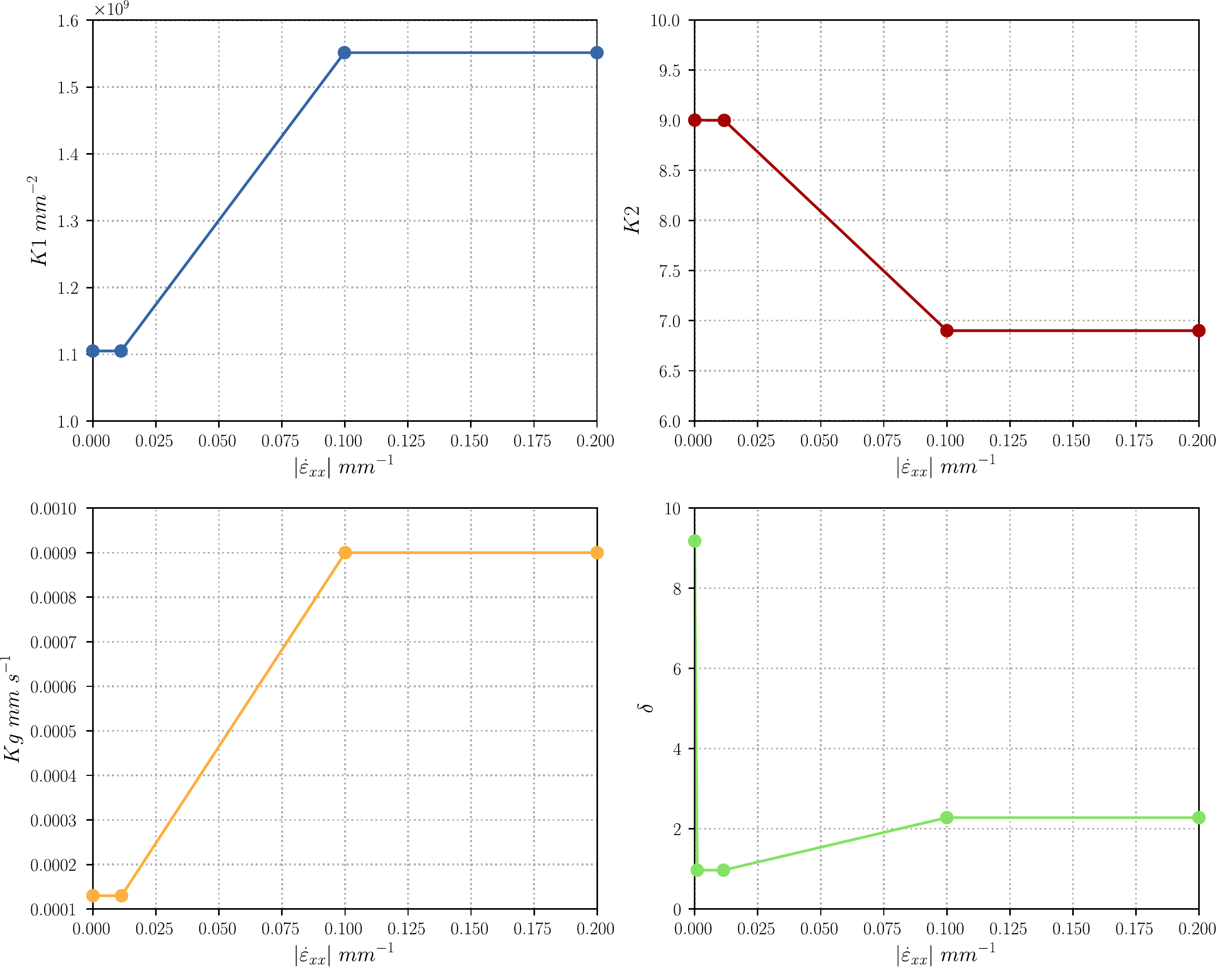}
\caption{Evolution of the parameters of table \ref{table:DataTable} in function of $|\dot{\varepsilon}_{xx}|$.}
\label{fig:DataFigure}
\end{figure}

\begin{figure}[!h]
\centering
\includegraphics[width=1.0\textwidth] {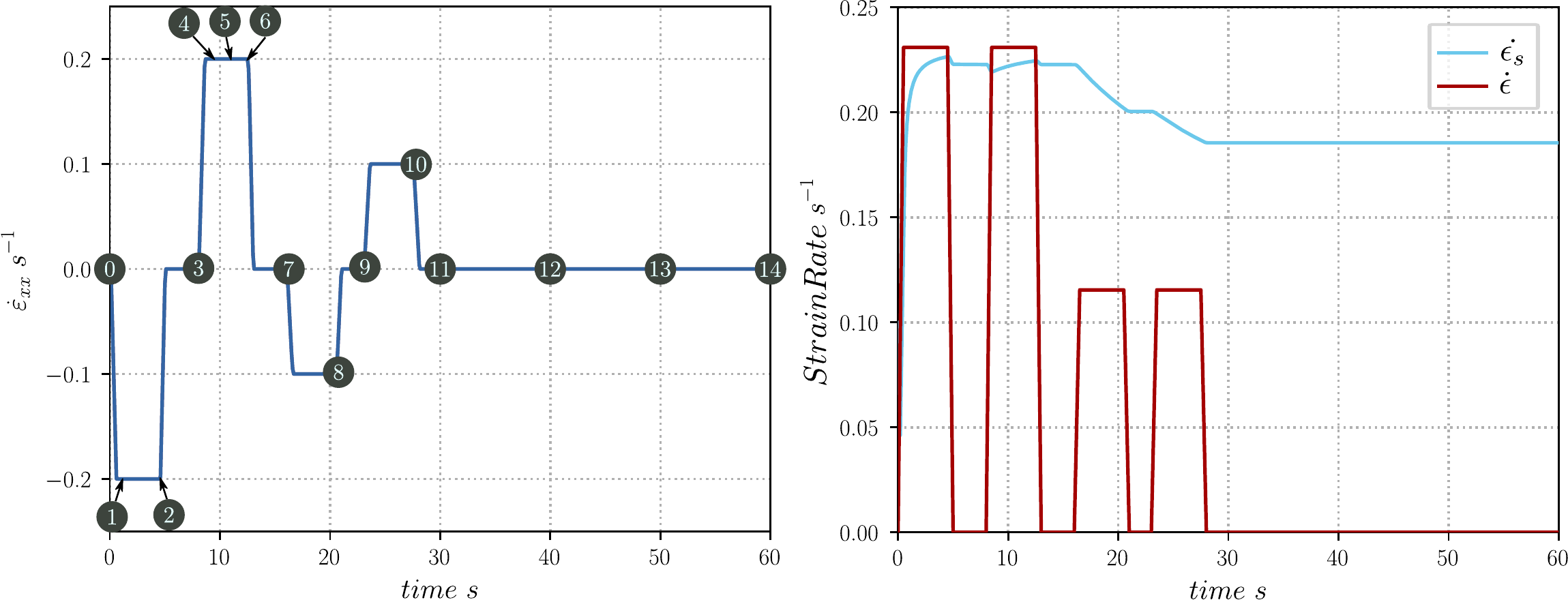}
\caption{Deformation loading strategy for the DRX and PDRX case: right: the computed values of the effective strain rates $\dot{\epsilon_s}$ and $\dot{\epsilon}$, left: the strain deformation component $\dot{\varepsilon}_{xx}$, where multiple markers have been drawn, corresponding to different states during the simulations.}
\label{fig:EffDefRate}
\end{figure}

Moreover, during PDRX ($|\dot{\varepsilon}_{xx}| = 0$), the parameter $\delta$ will take the value of $9.18$ following the findings in \cite{Maire2019}. Also, as explained in section \ref{sec:NucLaws}, during PDRX the parameter $\rho_c$ will be computed using the apparent effective strain rate $\dot{\epsilon_s}$ (see Eq. \ref{Eq:StaticEpsilonEquation} and Fig. \ref{fig:EffDefRate}.right) instead of the effective strain rate $\dot{\epsilon}$ (equals to 0 in this regime). Finally, outside the range of interpolation, the values are computed as follows: if $|\dot{\varepsilon}_{xx}|>0.1$ the values of all parameters will take the same values as for $|\dot{\varepsilon}_{xx}|=0.1~s^{-1}$, similarly, the same strategy applies when $|\dot{\varepsilon}_{xx}|<0.01$, using the values for $|\dot{\varepsilon}_{xx}|=0.01~s^{-1}$ (see Fig. \ref{fig:DataFigure} for an illustration of the values of $K1$, $K2$, $Kg$ and $\delta$ in function of $|\dot{\varepsilon}_{xx}|$). \\

Four cycles of deformation/coarsening will be applied as illustrated in Fig. \ref{fig:EffDefRate}, Fig. \ref{fig:EffDefRate}.right shows the computed values of the effective strain rates $\dot{\epsilon_s}$ and $\dot{\epsilon}$, while Fig. \ref{fig:EffDefRate}.left shows the strain deformation component $\dot{\varepsilon}_{xx}$, where multiple markers have been displayed, these markers correspond to different states along the simulation that will be useful when analyzing the results.\\

Statistical comparisons of the TRM model and the response obtained by a FE-LS approach presented in \cite{Bernacki2008, Cruz-Fabiano2014, Maire2017, Loge2008} will be given. This approach uses a more classic method of mesh adaptation during calculations where the interfaces are captured with an anisotropic non-conform local refined mesh. This methodology will be denoted in the following as the Anisotropic Meshing Adaptation (AMA) model.\\

A well known behavior of full field simulations of microstructural evolutions is that the reduced mobility ($\gamma M$ product) is classically impacted by the choice of the numerical method and is not only a universal physical parameter. In other words, the reduced mobility is a physical parameter that needs to be identified comparatively to experimental data. This identification may lead to different values depending generally of the numerical method used \cite{Villaret2020}. In \cite{Florez2020b}, the reduced mobility was adjusted in order to minimize the L2-difference between the mean grain size evolution curves considering TRM or AMA numerical strategies. Same methodology was used here in the global thermomechanical paths leading to an increase of 40\% in the optimal identified reduced mobility.\\

%In \cite{Florez2020b} we have adjusted the value of the mobility so the L2-Error between the curves of evolution of the mean grain size were minimized. Data from that study suggest that an increase on the value of the mobility parameter $M$ of $40.0\%$ is needed in order to match the behaviour of the TRM model to the AMA case.\\

Here, we have chosen the AMA case as a reference even though there is no way to know which model gives the most accurate response to the given physical problem; this choice on the other hand is given as an example of how the TRM model can indeed obtain similar responses to well established models in the field of microstructural evolutions.\\

Multiple microstructural states have been retrieved from the results given by the TRM model. This states are marked with numbers corresponding to the states of figure \ref{fig:EffDefRate}: \\

\begin{figure}[!h]
\centering
\includegraphics[width=1.0\textwidth] {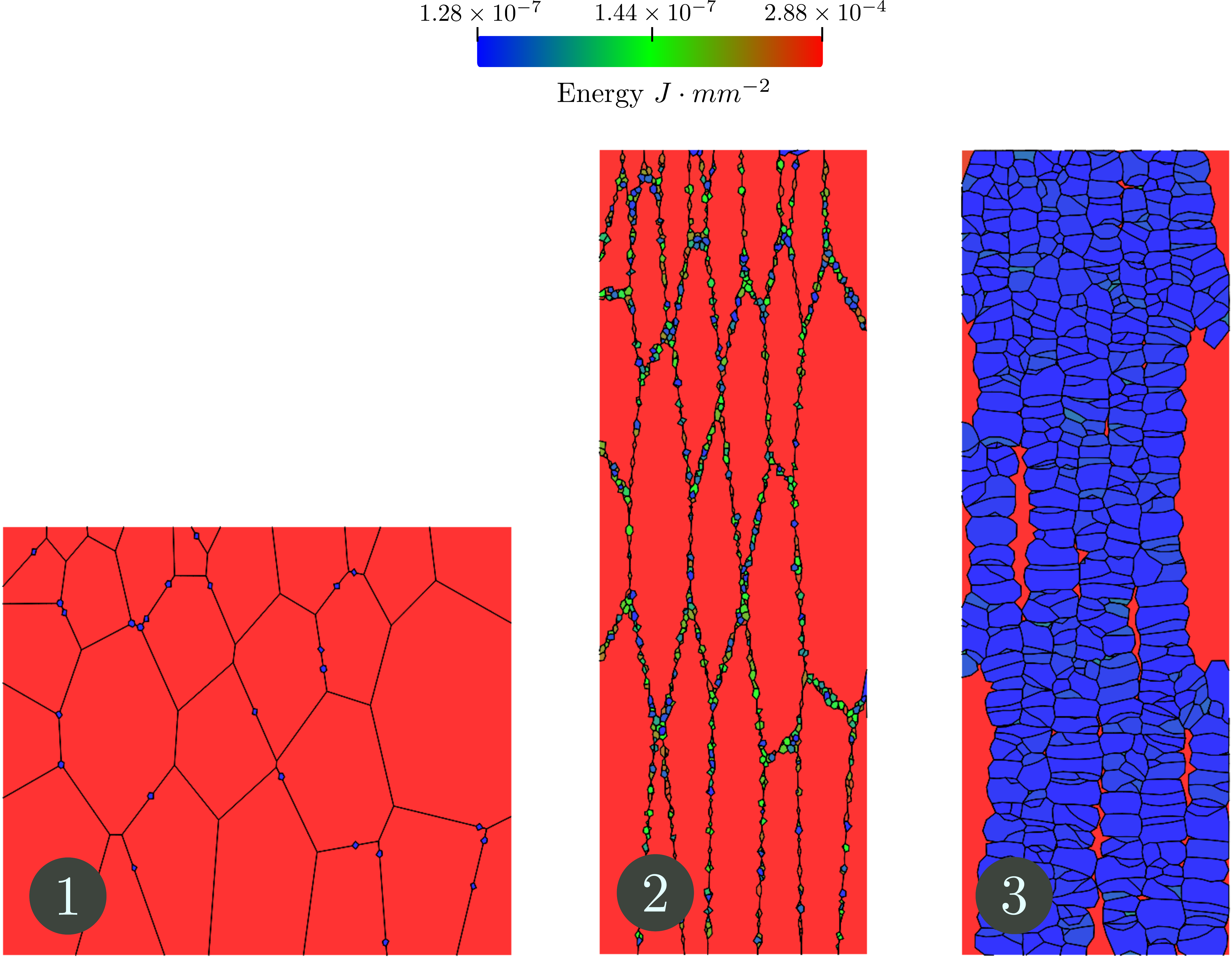}
\caption{States 1 to 3 (see figure \ref{fig:EffDefRate}) obtained with the TRM model. 1: the firsts nuclei appear, 2: end of the first stage of deformation, 2: end of the first grain coarsening stage.}
\label{fig:states1_3}
\end{figure}

States 1 to 3 are given in figure \ref{fig:states1_3}, here state 1 illustrates the apparition of the firsts nuclei in the positions where the dislocation density field reaches its value $\rho \geqslant \rho_c$. Then state 2 gives the end of the first stage of deformation where more nuclei have appeared, note that the value of the stored energy in some of the small grains is different from others, these grains have been present longer in the domain and consequently have been subjected to strain hardening, contrary to the nucleus that have appeared later, during or at the end of this deformation stage. Finally state 3 shows the end of the first grain coarsening stage where nuclei have been given time to growth as a product of the high difference in energy with their surroundings.\\

\begin{figure}[!h]
\centering
\includegraphics[width=1.0\textwidth] {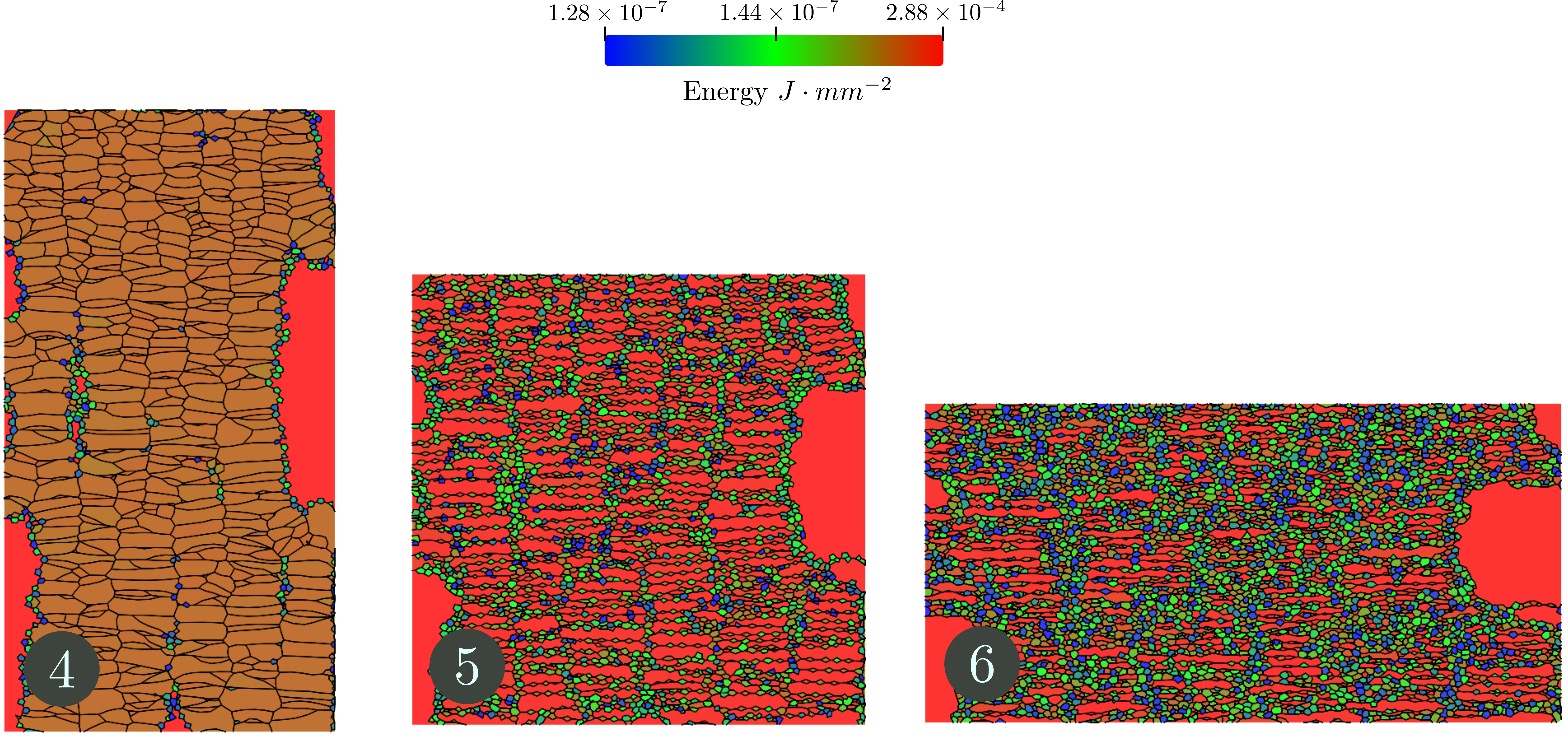}
\caption{States 4 to 6 (see figure \ref{fig:EffDefRate}) obtained with the TRM model. 4: the nuclei appear on the regions where $\rho>\rho_c$, 5: all the domain is now above the value of $\rho_c$ hence the nucleation occurs everywhere, 6: end of the second stage of deformation, here the maximum number of grains is reached (4250 grains).}
\label{fig:states4_6}
\end{figure}

States 4 to 6 are presented in Fig. \ref{fig:states4_6}. In stage 4 only a small percent of the domain have a dislocation density of at least $\rho_c$ and nucleation is restricted to these zones, contrary to stage 5, where a bigger part of the domain have reached the value of $\rho_c$, consequently new grains appear everywhere. Finally, the end of the second deformation stage is given in state 6 where the first peak of number of grains is reached (4250 grains).\\

\begin{figure}[!h]
\centering
\includegraphics[width=1.0\textwidth] {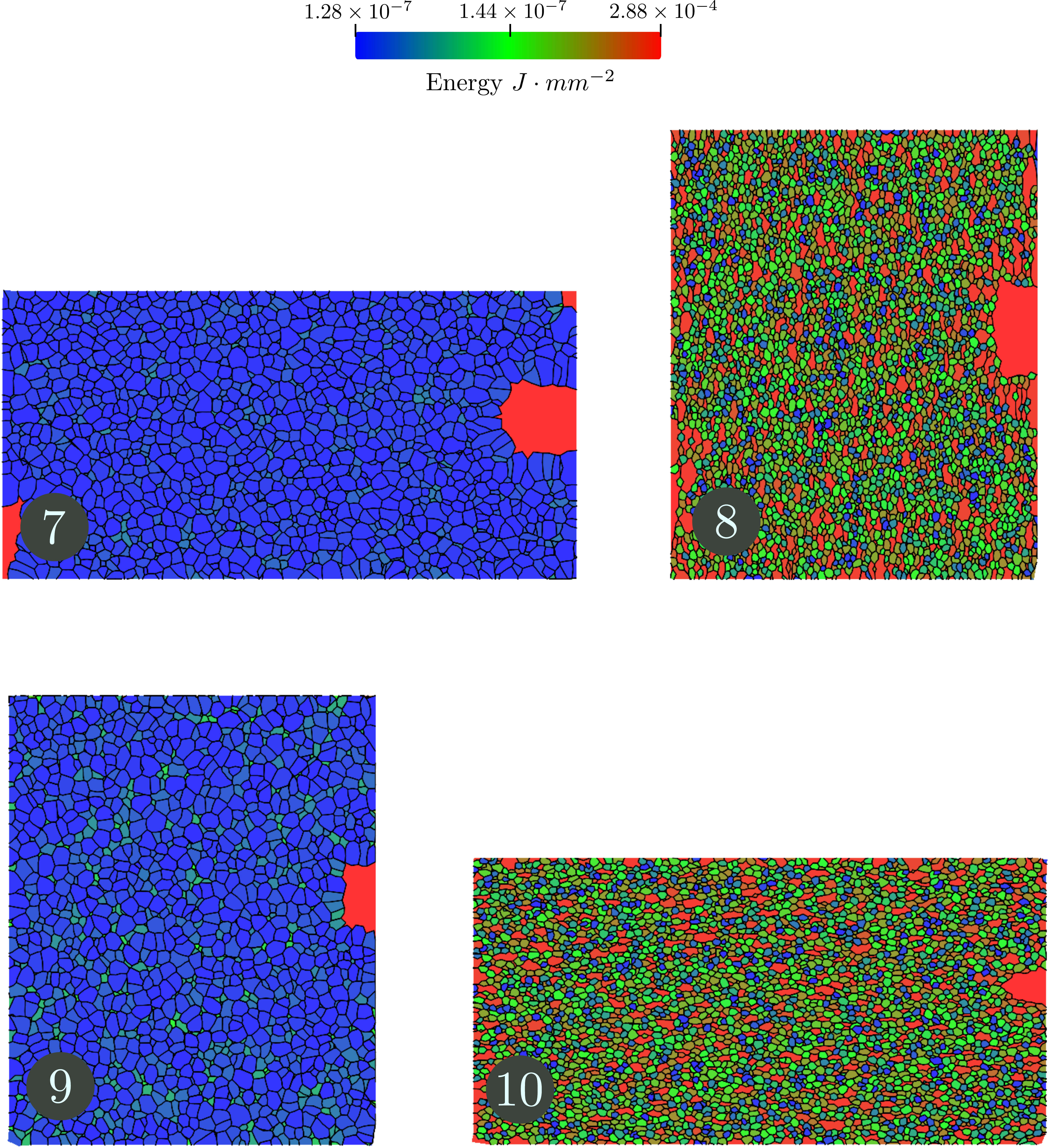}
\caption{States 7 to 10 (see figure \ref{fig:EffDefRate}) obtained with the TRM model. 7: end of the second grain coarsening stage, the number of grains drops very quickly given by the increase of the value of $\delta$ from 2.245 to 9.18 (its dynamic vs its static value) , 8: end of the third deformation stage, 9: end of the third grain coarsening stage, 10: end of the fourth deformation stage.}
\label{fig:states7_10}
\end{figure}

\begin{figure}[!h]
\centering
\includegraphics[width=1.0\textwidth] {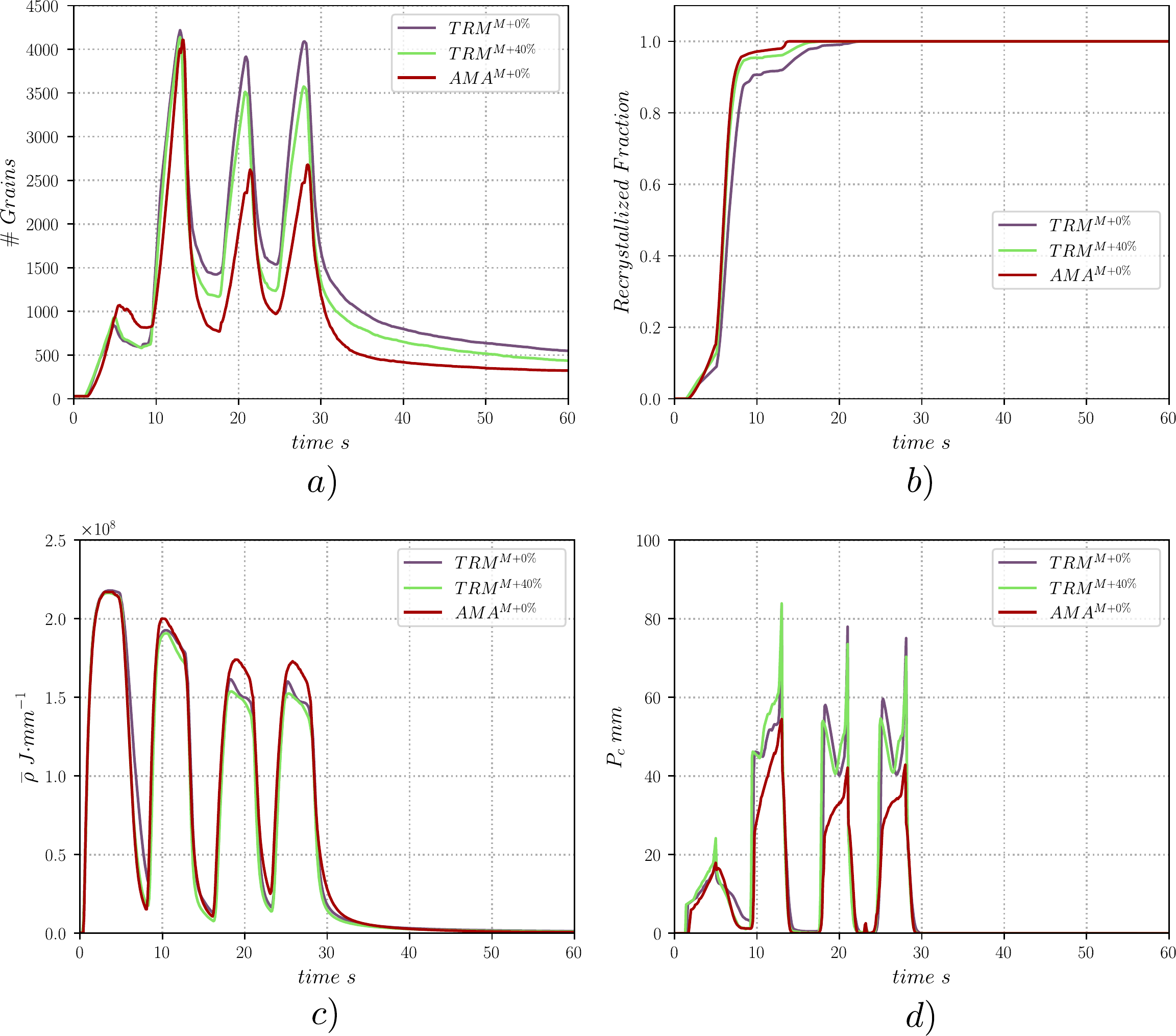}
\caption{different values as a function of time for the DRX and PDRX test case, a) Number of grains, b) Recrystallized fraction, c) Mean value of $\rho$ pondered by surface, and d) Critical perimeter for the computation of the nucleation rate in equation \ref{Eq:nucleationRateDrX}}
\label{fig:Statistics}
\end{figure}

States 7 to 10 are given in figure \ref{fig:states7_10}, these steps are representative of the ends of the third and fourth deformation/coarsening cycles, where during the deformation the nucleation process increases the number of grains while in the grain coarsening stages the high value of the parameter $\delta$ (2.245 to 9.18 its dynamic vs its static value) makes the grain number decrease rapidly (see figure \ref{fig:Statistics}.a) ) for the evolution of the number of grains).\\

\begin{figure}[!h]
\centering
\includegraphics[width=1.0\textwidth] {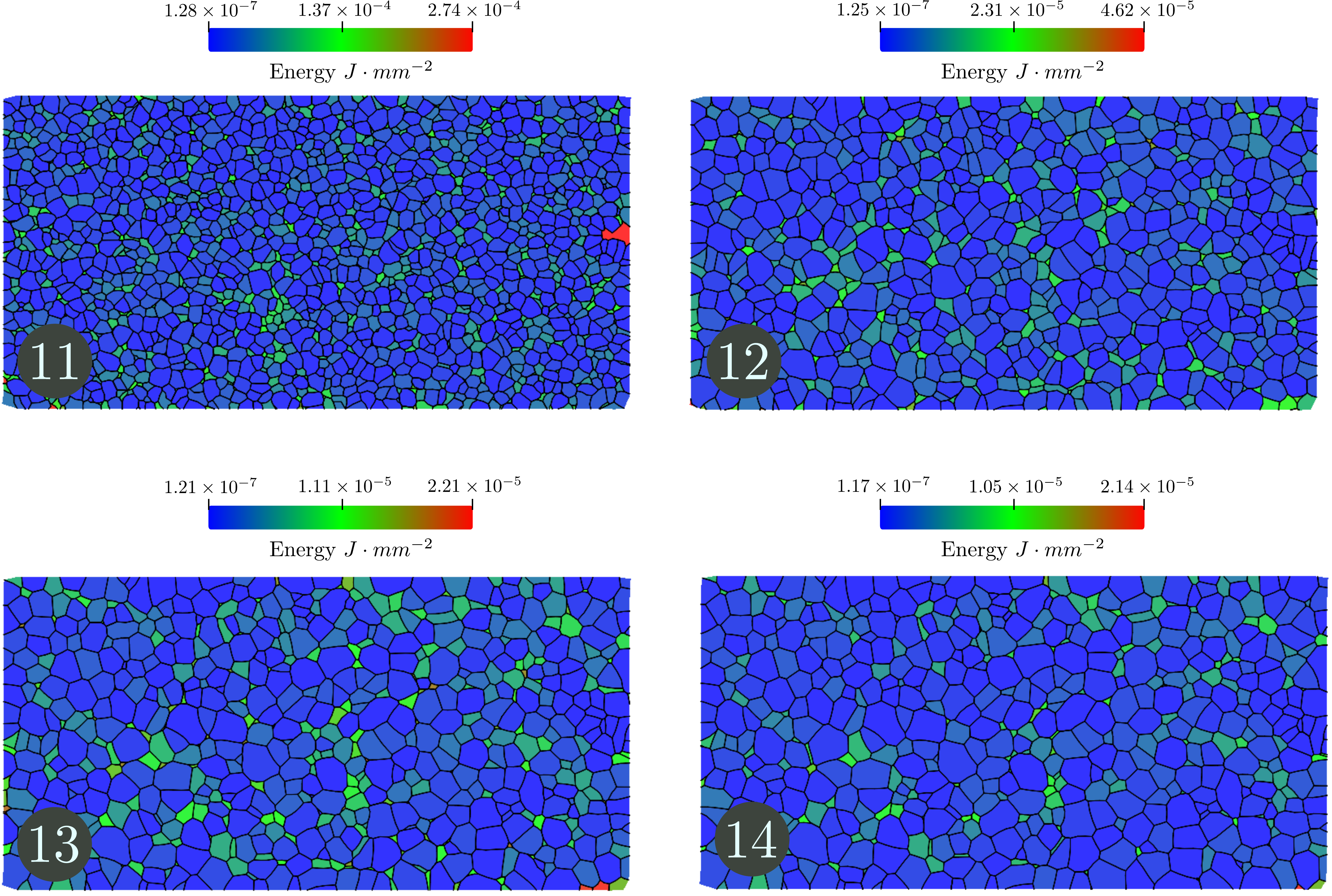}
\caption{States 11 to 14 (see figure \ref{fig:EffDefRate}) obtained with the TRM model. these state correspond to a value of time $t$ of 30, 40, 50 and 60 seconds respectively. }
\label{fig:states11_14}
\end{figure}

States 11 to 14 are provided in figure \ref{fig:states11_14}, 
these states correspond to a value of time $t$ of 30, 40, 50 and 60 seconds respectively. In this range of time no deformation is considered. Note how the limits of the scale in figure \ref{fig:states11_14} change as a product of the disappearance of high energetic grains and to the annihilation of dislocations simulated through equation \ref{Eq:annihilation}.\\

%Statistical values for the states 4 to 6 and 11 to 14 are given in figures \ref{fig:Histo4-6} and \ref{fig:Histo11-14} respectively, here the grain size distributions for the TRM model with an increase of $0\%$, $40\%$ and $80\%$ have been plotted along with the response given by the AMA case (simulations with a $20\%$, $60\%$ and $100\%$ have not been plotted so the data can be detailed.). Finally the evolution of the mean grain size are provided for all simulations  in figure \ref{fig:MeanSizeAndError}.left and its L2-Error to the AMA case is given in figure \ref{fig:MeanSizeAndError}.right\\

\begin{figure}[!h]
\centering
\includegraphics[width=1.0\textwidth] {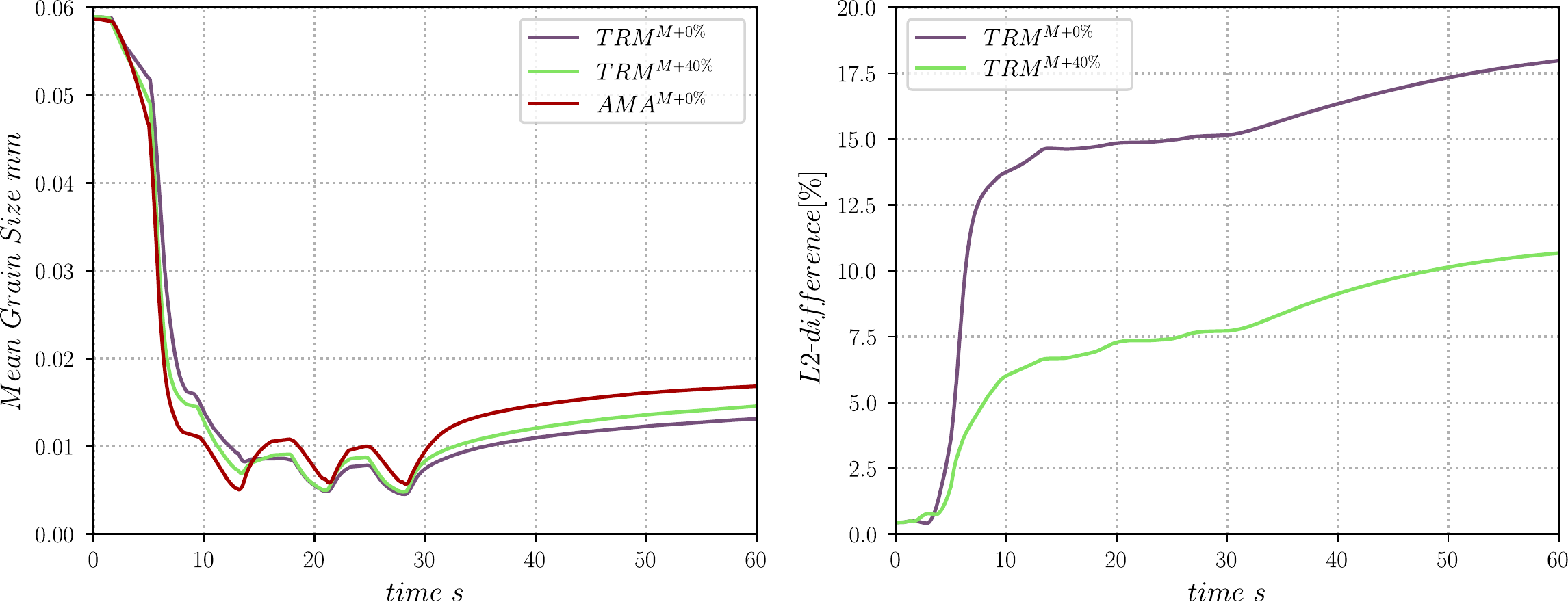}
\caption{Evolution of the Mean grain size (left) for the TRM model and the L2-difference with the AMA simulation (right).}
\label{fig:MeanSizeAndError}
\end{figure}

\begin{figure}[!h]
\centering
\includegraphics[width=1.0\textwidth] {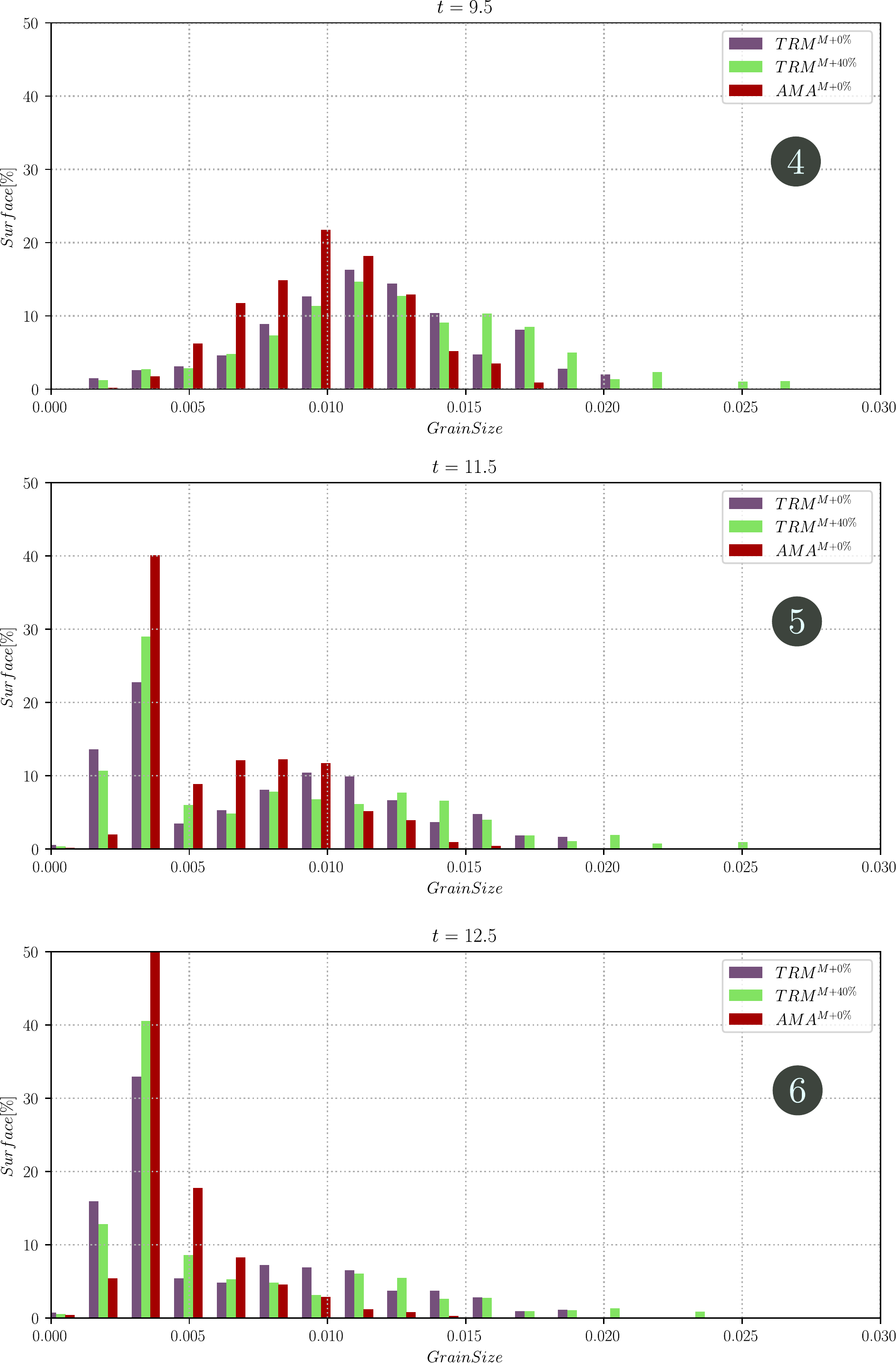}
\caption{Grain size distributions pondered in surface for the states 4 to 6 (example of a deformation Stage). A peek on the nucleus size can be observed.}
\label{fig:Histo4-6}
\end{figure}

\begin{figure}[!h]
\centering
\includegraphics[width=1.0\textwidth] {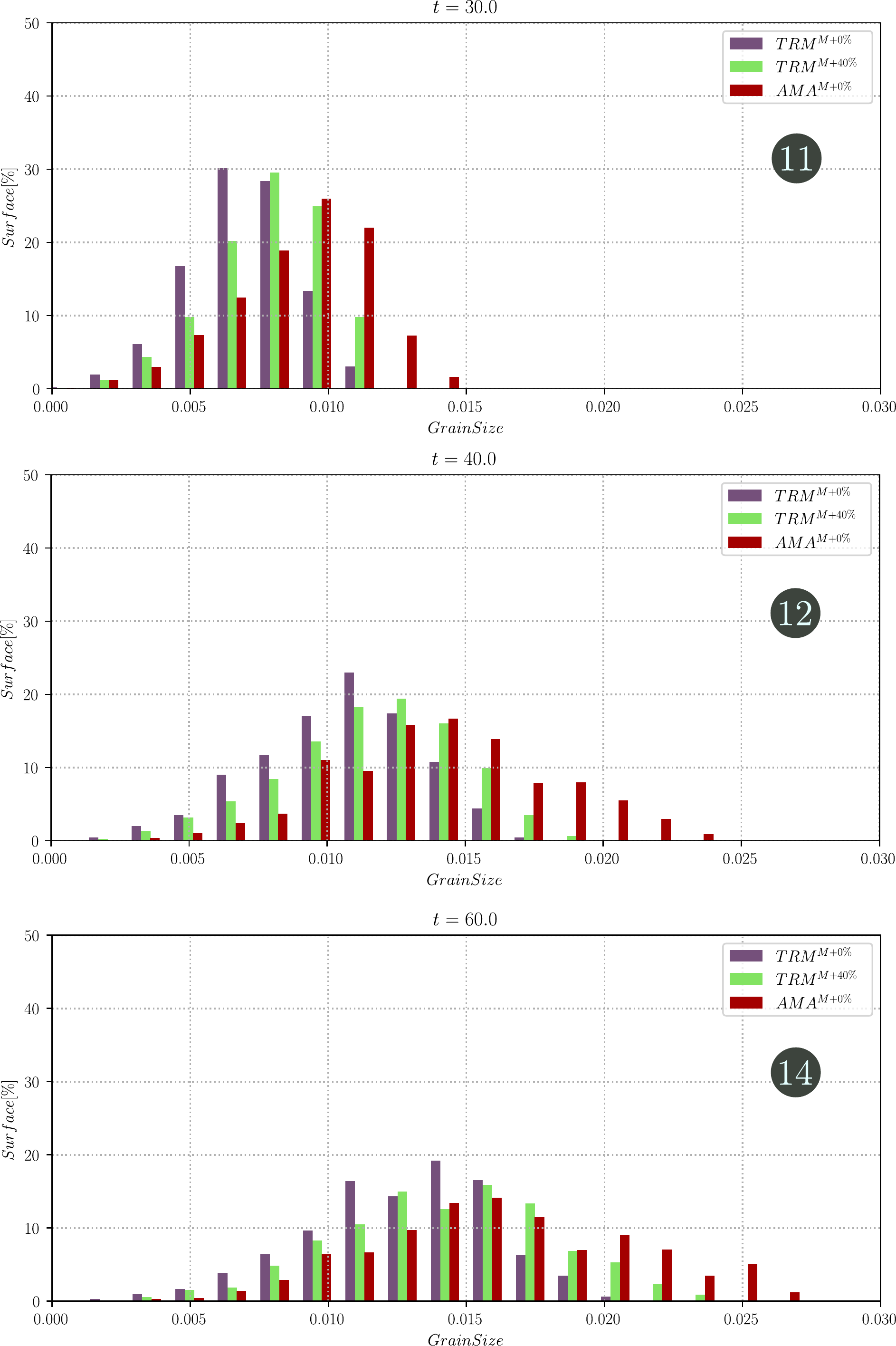}
\caption{Grain size distributions pondered in surface for the states 11, 13 and 14 (example of a grain coarsening stage). The values are distributed more evenly on the size range (x axis) as a product of the grain growth.}
\label{fig:Histo11-14}
\end{figure}

Statistical values for the states 4 to 6 and 11 to 14 are given in figures \ref{fig:Histo4-6} and \ref{fig:Histo11-14} respectively. The grain size distributions for the TRM model without a mobility increase and with a mobility increase of $40\%$ have been plotted along with the response given by the AMA case. Similarly the evolution of the mean grain size are provided for all simulations in figure \ref{fig:MeanSizeAndError}.left and the L2-difference to the AMA case is given in figure \ref{fig:MeanSizeAndError}.right\\

Finally, the evolution of some representative values are given in Fig. \ref{fig:Statistics}: the evolution of the number of grains, the recrystallized fraction, the mean value of $\rho$ pondered in surface ($\overline{\rho}$) and the total perimeter of the grains whose dislocation density is greater than $\rho_c$ ($P_c$) are provided.\\

%These results show that the TRM model approaches the behaviour of the AMA case when an increase of 80-100 $\%$ is made to the mobility $M$ value, contrary to our initial guess of $40\%$ obtained with the results of \cite{Florez2020b}. The computational cost for the different iterations of the TRM model is given in figure \ref{fig:CPU-Time}, where for the slower simulation the time needed for its completion was of 25 min and for the fastest of 20 min, compared to the time needed for the AMA case (-- min).

\begin{figure}[!h]
\centering
\includegraphics[width=0.7\textwidth] {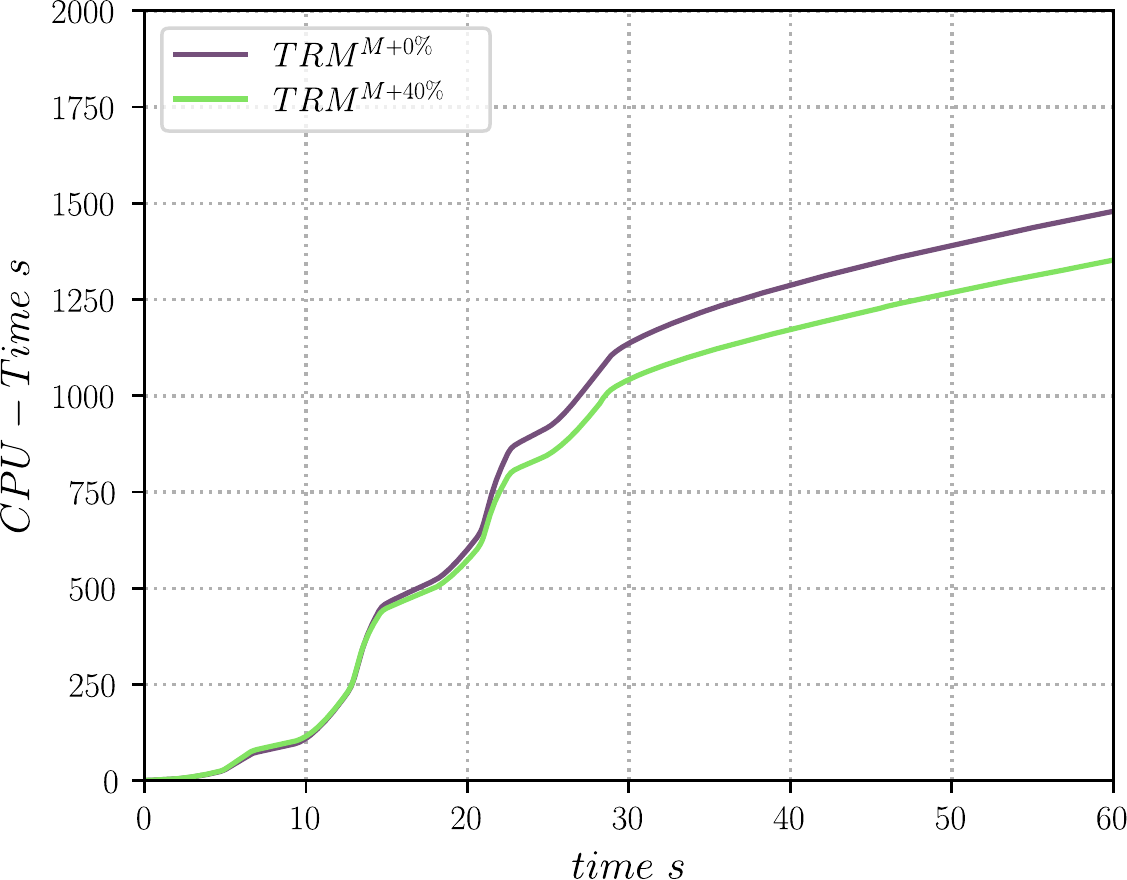}
\caption{CPU-time for the different simulations using the TRM model, the computational cost drops as the number of simulated grains decreases.}
\label{fig:CPU-Time}
\end{figure}

These results show a good agreement between the general behavior of the the TRM model and the behavior of the AMA simulation when an increase of 40$\%$ is considered to the reduced mobility $M\gamma$ value (following the findings in \cite{Florez2020b}). The computational cost for the different iterations of the TRM model is given in figure \ref{fig:CPU-Time}, where for the slower simulation the time needed for its completion was of 25 min and for the fastest of 20 min, compared to the time needed for the AMA case (4 hours and 38 min).

%% file: Conclusions.tex
\section{Discussion, conclusion and perspectives}\label{sec:conclusions}

In this article the TRM model presented in previous works in the context of isotropic grain growth by capillarity has been adapted in order to take into account bulk terms due to the stored energy during plastic deformation. This adaptation has made possible the integration of a recrystallization model to the TRM approach, for which a nucleation procedure has also been presented.\\

The algorithms presented in section \ref{sec:multJunctVelocity} and represented by Eq.  \ref{Eq:VelocityEquationeMP} for the computation of the velocity at multiple junctions, although intuitive have not been published before to the knowledge of the authors, only \cite{Hallberg2013} shows a similar (more indirect) approach in the context of vertex simulations.\\

Results for the circle test case and tripe junction case have demonstrated the high accuracy of the TRM model in the modeling of boundary migration due to capillarity and stored energy, where in the normal context (for typical grain boundaries and multiple junctions), an error no greater than $2\%$ was found. Also, the circle test case showed the typical behavior of a nucleus when subjected to a wide range of stored energy around its metastable point and helped define the safety factor $\omega$ used in Eq. \ref{Eq:MinimalRadiusNuclei}, defining the minimal radius to nucleate in the context of the TRM model.\\

%The bubble grain test allowed to give an insight of the current state of the TRM model and the wide variety of possibilities that this approach offers in the context of multi-domain simulations.\\

Finally, a DRX/PDRX test case was considered in order to test the recrystallization model provided in section \ref{sec:ReX} for 304L stainless steel at $1100$ $^{\circ}C$. A reference test case using the same ReX model but with a FE-LS strategy was also considered (AMA case) \cite{Bernacki2008, Cruz-Fabiano2014, Maire2017, Loge2008}. Following the findings in \cite{Florez2020b}, an optimal reduced mobility was calibrated to performed the tests (40$\%$ higher than the mobility used in the AMA case context). Results shows a very good agreement between the two models. Moreover the computational cost of the TRM model was lower being between 20 to 25 minutes against the 4 hours and 38 minutes needed for the AMA case for its completion.\\

Perspectives for the presented work and the TRM approach include the implementation of a model able to treat full anisotropic boundary properties, as well as the study of the in-grain gradients of stored energy. The 3D implementation of the TRM model will also be studied in future works.\\

\section*{Acknowledgments}
The authors thank the ArcelorMittal, ASCOMETAL, AUBERT \& DUVAL, CEA,  FRAMATOME, SAFRAN, TIMET, Constellium and TRANSVALOR companies and the ANR for their financial support through the DIGIMU consortium and ANR industrial Chair (Grant No. ANR-16-CHIN-0001).

\section*{Data availability}
The raw data required to reproduce these findings cannot be shared at this time as the data also forms part of an ongoing study. The processed data required to reproduce these findings cannot be shared at this time as the data also forms part of an ongoing study.